\documentclass[longauth]{aa}

\usepackage{hyperref}

\usepackage{xcolor}

\usepackage{graphicx}
\usepackage{siunitx}

\usepackage{txfonts}

\usepackage{euclid}

\newcommand{\SN}{\mathcal{S}}
\newcommand{\Ha}{\ha}
\newcommand{\Hb}{\hb}

\DeclareSIUnit{\erg}{erg}
\DeclareSIUnit{\counts}{counts}
\newcommand{\ergcms}{{\rm erg\,s^{-1}\,cm^{-2}}}
\DeclareSIUnit\angstrom{\protect \text {Å}}

\newcommand{\fastspec}{\texttt{FastSpec}\xspace}
\newcommand{\pypelid}{\texttt{Pypelid}\xspace}

\newcommand{\citelee}{Euclid Collaboration: Lee et al. (in prep.)\xspace}

\newcommand{\citeppassa}{(Euclid Collaboration: Passalacqua et al., in prep.)\xspace}

\begin{document}
\newcommand{\orcid}[1]{} 
\author{Euclid Collaboration: B.~R.~Granett\orcid{0000-0003-2694-9284}\corrauth{benjamin.granett@inaf.it}\inst{\ref{aff1}}
\and S.~de~la~Torre\inst{\ref{aff2}}
\and M.~Moresco\orcid{0000-0002-7616-7136}\inst{\ref{aff3},\ref{aff4}}
\and A.~Farina\orcid{0009-0000-3420-929X}\inst{\ref{aff1},\ref{aff5}}
\and M.~Bethermin\orcid{0000-0002-3915-2015}\inst{\ref{aff6}}
\and B.~Zabelle\orcid{0000-0002-7830-363X}\inst{\ref{aff7}}
\and S.~Bruton\orcid{0000-0002-6503-5218}\inst{\ref{aff8}}
\and K.~Markovic\orcid{0000-0001-6764-073X}\inst{\ref{aff9}}
\and I.~Risso\orcid{0000-0003-2525-7761}\inst{\ref{aff10},\ref{aff5},\ref{aff1}}
\and S.~Lee\orcid{0000-0002-8289-740X}\inst{\ref{aff9},\ref{aff11}}
\and K.~S.~McCarthy\orcid{0000-0001-6857-018X}\inst{\ref{aff9},\ref{aff12}}
\and O.~Ilbert\orcid{0000-0002-7303-4397}\inst{\ref{aff2}}
\and V.~Le~Brun\orcid{0000-0002-5027-1939}\inst{\ref{aff2}}
\and P.~Monaco\orcid{0000-0003-2083-7564}\inst{\ref{aff13},\ref{aff14},\ref{aff15},\ref{aff16}}
\and K.~V.~Nedkova\orcid{0000-0001-5294-8002}\inst{\ref{aff17}}
\and C.~Scarlata\orcid{0000-0002-9136-8876}\inst{\ref{aff7}}
\and L.~Guzzo\orcid{0000-0001-8264-5192}\inst{\ref{aff18},\ref{aff1},\ref{aff19}}
\and W.~J.~Percival\orcid{0000-0002-0644-5727}\inst{\ref{aff20},\ref{aff21},\ref{aff22}}
\and Y.~Wang\orcid{0000-0002-4749-2984}\inst{\ref{aff17}}
\and G.~Ca\~nas-Herrera\orcid{0000-0003-2796-2149}\inst{\ref{aff23}}
\and C.~Sirignano\orcid{0000-0002-0995-7146}\inst{\ref{aff24},\ref{aff25}}
\and B.~Altieri\orcid{0000-0003-3936-0284}\inst{\ref{aff26}}
\and S.~Andreon\orcid{0000-0002-2041-8784}\inst{\ref{aff1}}
\and N.~Auricchio\orcid{0000-0003-4444-8651}\inst{\ref{aff4}}
\and C.~Baccigalupi\orcid{0000-0002-8211-1630}\inst{\ref{aff16},\ref{aff14},\ref{aff15},\ref{aff27}}
\and M.~Baldi\orcid{0000-0003-4145-1943}\inst{\ref{aff28},\ref{aff4},\ref{aff29}}
\and S.~Bardelli\orcid{0000-0002-8900-0298}\inst{\ref{aff4}}
\and P.~Battaglia\orcid{0000-0002-7337-5909}\inst{\ref{aff4}}
\and A.~Biviano\orcid{0000-0002-0857-0732}\inst{\ref{aff14},\ref{aff16}}
\and M.~Bolzonella\orcid{0000-0003-3278-4607}\inst{\ref{aff4}}
\and E.~Branchini\orcid{0000-0002-0808-6908}\inst{\ref{aff10},\ref{aff5},\ref{aff1}}
\and M.~Brescia\orcid{0000-0001-9506-5680}\inst{\ref{aff30},\ref{aff31}}
\and J.~Brinchmann\orcid{0000-0003-4359-8797}\inst{\ref{aff32},\ref{aff33},\ref{aff34}}
\and S.~Camera\orcid{0000-0003-3399-3574}\inst{\ref{aff35},\ref{aff36},\ref{aff37}}
\and V.~Capobianco\orcid{0000-0002-3309-7692}\inst{\ref{aff37}}
\and C.~Carbone\orcid{0000-0003-0125-3563}\inst{\ref{aff38}}
\and J.~Carretero\orcid{0000-0002-3130-0204}\inst{\ref{aff39},\ref{aff40}}
\and M.~Castellano\orcid{0000-0001-9875-8263}\inst{\ref{aff41}}
\and G.~Castignani\orcid{0000-0001-6831-0687}\inst{\ref{aff4}}
\and S.~Cavuoti\orcid{0000-0002-3787-4196}\inst{\ref{aff31},\ref{aff42}}
\and P.-Y.~Chabaud\inst{\ref{aff2}}
\and A.~Cimatti\inst{\ref{aff43}}
\and C.~Colodro-Conde\inst{\ref{aff44}}
\and G.~Congedo\orcid{0000-0003-2508-0046}\inst{\ref{aff45}}
\and L.~Conversi\orcid{0000-0002-6710-8476}\inst{\ref{aff46},\ref{aff26}}
\and Y.~Copin\orcid{0000-0002-5317-7518}\inst{\ref{aff47}}
\and A.~Costille\inst{\ref{aff2}}
\and F.~Courbin\orcid{0000-0003-0758-6510}\inst{\ref{aff48},\ref{aff49},\ref{aff50}}
\and H.~M.~Courtois\orcid{0000-0003-0509-1776}\inst{\ref{aff51}}
\and H.~Degaudenzi\orcid{0000-0002-5887-6799}\inst{\ref{aff52}}
\and G.~De~Lucia\orcid{0000-0002-6220-9104}\inst{\ref{aff14}}
\and H.~Dole\orcid{0000-0002-9767-3839}\inst{\ref{aff53}}
\and F.~Dubath\orcid{0000-0002-6533-2810}\inst{\ref{aff52}}
\and X.~Dupac\inst{\ref{aff26}}
\and S.~Dusini\orcid{0000-0002-1128-0664}\inst{\ref{aff25}}
\and S.~Escoffier\orcid{0000-0002-2847-7498}\inst{\ref{aff54}}
\and M.~Farina\orcid{0000-0002-3089-7846}\inst{\ref{aff55}}
\and R.~Farinelli\inst{\ref{aff4}}
\and F.~Faustini\orcid{0000-0001-6274-5145}\inst{\ref{aff41}}
\and S.~Ferriol\inst{\ref{aff47}}
\and F.~Finelli\orcid{0000-0002-6694-3269}\inst{\ref{aff4},\ref{aff56}}
\and P.~Fosalba\orcid{0000-0002-1510-5214}\inst{\ref{aff57},\ref{aff58}}
\and S.~Fotopoulou\orcid{0000-0002-9686-254X}\inst{\ref{aff59}}
\and N.~Fourmanoit\orcid{0009-0005-6816-6925}\inst{\ref{aff54}}
\and M.~Frailis\orcid{0000-0002-7400-2135}\inst{\ref{aff14}}
\and E.~Franceschi\orcid{0000-0002-0585-6591}\inst{\ref{aff4}}
\and L.~Gabarra\orcid{0000-0002-8486-8856}\inst{\ref{aff60}}
\and S.~Galeotta\orcid{0000-0002-3748-5115}\inst{\ref{aff14}}
\and K.~George\orcid{0000-0002-1734-8455}\inst{\ref{aff61}}
\and B.~Gillis\orcid{0000-0002-4478-1270}\inst{\ref{aff45}}
\and C.~Giocoli\orcid{0000-0002-9590-7961}\inst{\ref{aff4},\ref{aff29}}
\and J.~Gracia-Carpio\orcid{0000-0003-4689-3134}\inst{\ref{aff62}}
\and A.~Grazian\orcid{0000-0002-5688-0663}\inst{\ref{aff63}}
\and F.~Grupp\inst{\ref{aff62},\ref{aff64}}
\and S.~V.~H.~Haugan\orcid{0000-0001-9648-7260}\inst{\ref{aff65}}
\and S.~Hemmati\orcid{0000-0003-2226-5395}\inst{\ref{aff17}}
\and W.~Holmes\orcid{0009-0007-8554-4646}\inst{\ref{aff9}}
\and F.~Hormuth\inst{\ref{aff66}}
\and A.~Hornstrup\orcid{0000-0002-3363-0936}\inst{\ref{aff67},\ref{aff68}}
\and K.~Jahnke\orcid{0000-0003-3804-2137}\inst{\ref{aff69}}
\and M.~Jhabvala\inst{\ref{aff70}}
\and B.~Joachimi\orcid{0000-0001-7494-1303}\inst{\ref{aff71}}
\and S.~Kermiche\orcid{0000-0002-0302-5735}\inst{\ref{aff54}}
\and A.~Kiessling\orcid{0000-0002-2590-1273}\inst{\ref{aff9}}
\and B.~Kubik\orcid{0009-0006-5823-4880}\inst{\ref{aff47}}
\and M.~K\"ummel\orcid{0000-0003-2791-2117}\inst{\ref{aff64}}
\and M.~Kunz\orcid{0000-0002-3052-7394}\inst{\ref{aff72}}
\and H.~Kurki-Suonio\orcid{0000-0002-4618-3063}\inst{\ref{aff73},\ref{aff74}}
\and R.~Laureijs\inst{\ref{aff75}}
\and A.~M.~C.~Le~Brun\orcid{0000-0002-0936-4594}\inst{\ref{aff76}}
\and S.~Ligori\orcid{0000-0003-4172-4606}\inst{\ref{aff37}}
\and P.~B.~Lilje\orcid{0000-0003-4324-7794}\inst{\ref{aff65}}
\and V.~Lindholm\orcid{0000-0003-2317-5471}\inst{\ref{aff73},\ref{aff74}}
\and I.~Lloro\orcid{0000-0001-5966-1434}\inst{\ref{aff77}}
\and M.~Magliocchetti\orcid{0000-0001-9158-4838}\inst{\ref{aff55}}
\and G.~Mainetti\orcid{0000-0003-2384-2377}\inst{\ref{aff78}}
\and O.~Mansutti\orcid{0000-0001-5758-4658}\inst{\ref{aff14}}
\and O.~Marggraf\orcid{0000-0001-7242-3852}\inst{\ref{aff79}}
\and M.~Martinelli\orcid{0000-0002-6943-7732}\inst{\ref{aff41},\ref{aff80}}
\and N.~Martinet\orcid{0000-0003-2786-7790}\inst{\ref{aff2}}
\and F.~Marulli\orcid{0000-0002-8850-0303}\inst{\ref{aff3},\ref{aff4},\ref{aff29}}
\and R.~J.~Massey\orcid{0000-0002-6085-3780}\inst{\ref{aff81}}
\and N.~Mauri\orcid{0000-0001-8196-1548}\inst{\ref{aff43},\ref{aff29}}
\and E.~Medinaceli\orcid{0000-0002-4040-7783}\inst{\ref{aff4}}
\and M.~Meneghetti\orcid{0000-0003-1225-7084}\inst{\ref{aff4},\ref{aff29}}
\and E.~Merlin\orcid{0000-0001-6870-8900}\inst{\ref{aff63}}
\and G.~Meylan\orcid{0000-0001-6503-0209}\inst{\ref{aff82}}
\and A.~Mora\orcid{0000-0002-1922-8529}\inst{\ref{aff83}}
\and C.~Moretti\orcid{0000-0003-3314-8936}\inst{\ref{aff14},\ref{aff16},\ref{aff15}}
\and L.~Moscardini\orcid{0000-0002-3473-6716}\inst{\ref{aff3},\ref{aff4},\ref{aff29}}
\and R.~Nakajima\orcid{0009-0009-1213-7040}\inst{\ref{aff79}}
\and C.~Neissner\orcid{0000-0001-8524-4968}\inst{\ref{aff84},\ref{aff40}}
\and S.-M.~Niemi\orcid{0009-0005-0247-0086}\inst{\ref{aff85}}
\and J.~W.~Nightingale\orcid{0000-0002-8987-7401}\inst{\ref{aff86}}
\and C.~Padilla\orcid{0000-0001-7951-0166}\inst{\ref{aff84}}
\and S.~Paltani\orcid{0000-0002-8108-9179}\inst{\ref{aff52}}
\and F.~Pasian\orcid{0000-0002-4869-3227}\inst{\ref{aff14}}
\and J.~A.~Peacock\orcid{0000-0002-1168-8299}\inst{\ref{aff45}}
\and V.~Pettorino\orcid{0000-0002-4203-9320}\inst{\ref{aff85}}
\and A.~Pezzotta\orcid{0000-0003-0726-2268}\inst{\ref{aff1}}
\and G.~Polenta\orcid{0000-0003-4067-9196}\inst{\ref{aff87}}
\and M.~Poncet\inst{\ref{aff88}}
\and L.~A.~Popa\inst{\ref{aff89}}
\and L.~Pozzetti\orcid{0000-0001-7085-0412}\inst{\ref{aff4}}
\and F.~Raison\orcid{0000-0002-7819-6918}\inst{\ref{aff62}}
\and A.~Renzi\orcid{0000-0001-9856-1970}\inst{\ref{aff24},\ref{aff25},\ref{aff4}}
\and J.~Rhodes\orcid{0000-0002-4485-8549}\inst{\ref{aff9}}
\and G.~Riccio\inst{\ref{aff31}}
\and F.~Rizzo\orcid{0000-0002-9407-585X}\inst{\ref{aff14}}
\and E.~Romelli\orcid{0000-0003-3069-9222}\inst{\ref{aff14}}
\and M.~Roncarelli\orcid{0000-0001-9587-7822}\inst{\ref{aff4}}
\and B.~Rusholme\orcid{0000-0001-7648-4142}\inst{\ref{aff17}}
\and R.~Saglia\orcid{0000-0003-0378-7032}\inst{\ref{aff64},\ref{aff62}}
\and Z.~Sakr\orcid{0000-0002-4823-3757}\inst{\ref{aff90},\ref{aff91},\ref{aff92}}
\and A.~G.~S\'anchez\orcid{0000-0003-1198-831X}\inst{\ref{aff62}}
\and D.~Sapone\orcid{0000-0001-7089-4503}\inst{\ref{aff93}}
\and B.~Sartoris\orcid{0000-0003-1337-5269}\inst{\ref{aff62},\ref{aff14}}
\and P.~Schneider\orcid{0000-0001-8561-2679}\inst{\ref{aff79}}
\and T.~Schrabback\orcid{0000-0002-6987-7834}\inst{\ref{aff94}}
\and M.~Scodeggio\inst{\ref{aff38}}
\and A.~Secroun\orcid{0000-0003-0505-3710}\inst{\ref{aff54}}
\and E.~Sihvola\orcid{0000-0003-1804-7715}\inst{\ref{aff95}}
\and G.~Sirri\orcid{0000-0003-2626-2853}\inst{\ref{aff29}}
\and L.~Stanco\orcid{0000-0002-9706-5104}\inst{\ref{aff25}}
\and J.-L.~Starck\orcid{0000-0003-2177-7794}\inst{\ref{aff96}}
\and C.~Surace\orcid{0000-0003-2592-0113}\inst{\ref{aff2}}
\and P.~Tallada-Cresp\'{i}\orcid{0000-0002-1336-8328}\inst{\ref{aff39},\ref{aff40}}
\and D.~Tavagnacco\orcid{0000-0001-7475-9894}\inst{\ref{aff14}}
\and A.~N.~Taylor\inst{\ref{aff45}}
\and I.~Tereno\orcid{0000-0002-4537-6218}\inst{\ref{aff97},\ref{aff98}}
\and N.~Tessore\orcid{0000-0002-9696-7931}\inst{\ref{aff99}}
\and S.~Toft\orcid{0000-0003-3631-7176}\inst{\ref{aff100},\ref{aff101}}
\and R.~Toledo-Moreo\orcid{0000-0002-2997-4859}\inst{\ref{aff102},\ref{aff103}}
\and F.~Torradeflot\orcid{0000-0003-1160-1517}\inst{\ref{aff40},\ref{aff39}}
\and I.~Tutusaus\orcid{0000-0002-3199-0399}\inst{\ref{aff58},\ref{aff57},\ref{aff91}}
\and J.~Valiviita\orcid{0000-0001-6225-3693}\inst{\ref{aff73},\ref{aff74}}
\and T.~Vassallo\orcid{0000-0001-6512-6358}\inst{\ref{aff14},\ref{aff61}}
\and A.~Veropalumbo\orcid{0000-0003-2387-1194}\inst{\ref{aff1},\ref{aff5},\ref{aff10}}
\and D.~Vibert\orcid{0009-0008-0607-631X}\inst{\ref{aff2}}
\and J.~Weller\orcid{0000-0002-8282-2010}\inst{\ref{aff64},\ref{aff62}}
\and A.~Zacchei\orcid{0000-0003-0396-1192}\inst{\ref{aff14},\ref{aff16}}
\and G.~Zamorani\orcid{0000-0002-2318-301X}\inst{\ref{aff4}}
\and F.~M.~Zerbi\orcid{0000-0002-9996-973X}\inst{\ref{aff1}}
\and I.~A.~Zinchenko\orcid{0000-0002-2944-2449}\inst{\ref{aff104}}
\and E.~Zucca\orcid{0000-0002-5845-8132}\inst{\ref{aff4}}
\and M.~Ballardini\orcid{0000-0003-4481-3559}\inst{\ref{aff105},\ref{aff106},\ref{aff4}}
\and P.~Bergamini\orcid{0000-0003-1383-9414}\inst{\ref{aff4}}
\and A.~Boucaud\orcid{0000-0001-7387-2633}\inst{\ref{aff107}}
\and E.~Bozzo\orcid{0000-0002-8201-1525}\inst{\ref{aff52}}
\and C.~Burigana\orcid{0000-0002-3005-5796}\inst{\ref{aff108},\ref{aff56}}
\and R.~Cabanac\orcid{0000-0001-6679-2600}\inst{\ref{aff91}}
\and M.~Calabrese\orcid{0000-0002-2637-2422}\inst{\ref{aff109},\ref{aff38}}
\and A.~Cappi\inst{\ref{aff110},\ref{aff4}}
\and F.~Caro\orcid{0009-0003-1053-0507}\inst{\ref{aff41}}
\and T.~Castro\orcid{0000-0002-6292-3228}\inst{\ref{aff111},\ref{aff14},\ref{aff16}}
\and G.~Daste\inst{\ref{aff2}}
\and J.~A.~Escartin~Vigo\inst{\ref{aff62}}
\and J.~Garc\'ia-Bellido\orcid{0000-0002-9370-8360}\inst{\ref{aff90}}
\and T.~Gasparetto\orcid{0000-0002-7913-4866}\inst{\ref{aff41}}
\and V.~Gautard\inst{\ref{aff112}}
\and J.~Macias-Perez\orcid{0000-0002-5385-2763}\inst{\ref{aff113}}
\and R.~Maoli\orcid{0000-0002-6065-3025}\inst{\ref{aff114},\ref{aff41}}
\and R.~B.~Metcalf\orcid{0000-0003-3167-2574}\inst{\ref{aff3},\ref{aff4}}
\and A.~A.~Nucita\inst{\ref{aff115},\ref{aff116},\ref{aff117}}
\and M.~P\"ontinen\orcid{0000-0001-5442-2530}\inst{\ref{aff73}}
\and E.~Sarpa\orcid{0000-0002-1256-655X}\inst{\ref{aff14}}
\and V.~Scottez\orcid{0009-0008-3864-940X}\inst{\ref{aff118},\ref{aff119}}
\and M.~Sereno\orcid{0000-0003-0302-0325}\inst{\ref{aff4},\ref{aff29}}
\and M.~Tenti\orcid{0000-0002-4254-5901}\inst{\ref{aff29}}
\and M.~Tucci\inst{\ref{aff52}}
\and M.~Viel\orcid{0000-0002-2642-5707}\inst{\ref{aff16},\ref{aff14},\ref{aff27},\ref{aff15},\ref{aff120}}
\and M.~Wiesmann\orcid{0009-0000-8199-5860}\inst{\ref{aff65}}
\and J.~A.~Acevedo~Barroso\orcid{0000-0002-9654-1711}\inst{\ref{aff9}}
\and Y.~Akrami\orcid{0000-0002-2407-7956}\inst{\ref{aff90},\ref{aff121}}
\and I.~T.~Andika\orcid{0000-0001-6102-9526}\inst{\ref{aff64}}
\and S.~Anselmi\orcid{0000-0002-3579-9583}\inst{\ref{aff25},\ref{aff24},\ref{aff122}}
\and M.~Archidiacono\orcid{0000-0003-4952-9012}\inst{\ref{aff18},\ref{aff19}}
\and G.~Aric\`o\orcid{0000-0002-2802-2928}\inst{\ref{aff29}}
\and F.~Atrio-Barandela\orcid{0000-0002-2130-2513}\inst{\ref{aff123}}
\and S.~Avila\orcid{0000-0001-5043-3662}\inst{\ref{aff39}}
\and M.~Baes\orcid{0000-0002-3930-2757}\inst{\ref{aff124}}
\and L.~Bazzanini\orcid{0000-0003-0727-0137}\inst{\ref{aff105},\ref{aff4}}
\and J.~Bel\orcid{0009-0006-7837-1866}\inst{\ref{aff125}}
\and D.~Bertacca\orcid{0000-0002-2490-7139}\inst{\ref{aff24},\ref{aff63},\ref{aff25}}
\and F.~Beutler\orcid{0000-0003-0467-5438}\inst{\ref{aff45}}
\and A.~Blanchard\orcid{0000-0001-8555-9003}\inst{\ref{aff91}}
\and L.~Blot\orcid{0000-0002-9622-7167}\inst{\ref{aff126},\ref{aff76}}
\and M.~Bonici\orcid{0000-0002-8430-126X}\inst{\ref{aff20},\ref{aff38}}
\and M.~L.~Brown\orcid{0000-0002-0370-8077}\inst{\ref{aff127}}
\and A.~Calabro\orcid{0000-0003-2536-1614}\inst{\ref{aff41}}
\and B.~Camacho~Quevedo\orcid{0000-0002-8789-4232}\inst{\ref{aff16},\ref{aff27},\ref{aff14}}
\and P.~Carrilho\orcid{0000-0003-1339-0194}\inst{\ref{aff128}}
\and C.~S.~Carvalho\inst{\ref{aff98}}
\and F.~Cogato\orcid{0000-0003-4632-6113}\inst{\ref{aff3},\ref{aff4}}
\and T.~E.~Collett\orcid{0000-0001-5564-3140}\inst{\ref{aff129}}
\and S.~Conseil\orcid{0000-0002-3657-4191}\inst{\ref{aff47}}
\and S.~Contarini\orcid{0000-0002-9843-723X}\inst{\ref{aff62}}
\and A.~R.~Cooray\orcid{0000-0002-3892-0190}\inst{\ref{aff130}}
\and P.~Corcho-Caballero\orcid{0000-0001-6327-7080}\inst{\ref{aff75}}
\and B.~Csizi\orcid{0000-0003-3227-6581}\inst{\ref{aff94}}
\and O.~Cucciati\orcid{0000-0002-9336-7551}\inst{\ref{aff4}}
\and S.~Davini\orcid{0000-0003-3269-1718}\inst{\ref{aff5}}
\and T.~de~Boer\orcid{0000-0001-5486-2747}\inst{\ref{aff131}}
\and F.~De~Paolis\orcid{0000-0001-6460-7563}\inst{\ref{aff115},\ref{aff116},\ref{aff117}}
\and G.~Desprez\orcid{0000-0001-8325-1742}\inst{\ref{aff75}}
\and A.~D\'iaz-S\'anchez\orcid{0000-0003-0748-4768}\inst{\ref{aff132}}
\and S.~Di~Domizio\orcid{0000-0003-2863-5895}\inst{\ref{aff10},\ref{aff5}}
\and J.~M.~Diego\orcid{0000-0001-9065-3926}\inst{\ref{aff133}}
\and V.~Duret\orcid{0009-0009-0383-4960}\inst{\ref{aff54}}
\and M.~Y.~Elkhashab\orcid{0000-0001-9306-2603}\inst{\ref{aff14},\ref{aff15},\ref{aff13},\ref{aff16}}
\and A.~Enia\orcid{0000-0002-0200-2857}\inst{\ref{aff4}}
\and Y.~Fang\orcid{0000-0002-0334-6950}\inst{\ref{aff64}}
\and A.~Finoguenov\orcid{0000-0002-4606-5403}\inst{\ref{aff73}}
\and A.~Franco\orcid{0000-0002-4761-366X}\inst{\ref{aff115},\ref{aff116},\ref{aff117}}
\and Y.~Fu\orcid{0000-0002-0759-0504}\inst{\ref{aff23},\ref{aff75}}
\and K.~Ganga\orcid{0000-0001-8159-8208}\inst{\ref{aff107}}
\and E.~Gaztanaga\orcid{0000-0001-9632-0815}\inst{\ref{aff58},\ref{aff57},\ref{aff129}}
\and Z.~Ghaffari\orcid{0000-0002-6467-8078}\inst{\ref{aff14},\ref{aff16}}
\and F.~Giacomini\orcid{0000-0002-3129-2814}\inst{\ref{aff29}}
\and F.~Gianotti\orcid{0000-0003-4666-119X}\inst{\ref{aff4}}
\and G.~Gozaliasl\orcid{0000-0002-0236-919X}\inst{\ref{aff134},\ref{aff73}}
\and A.~Gruppuso\orcid{0000-0001-9272-5292}\inst{\ref{aff4},\ref{aff29}}
\and M.~Guidi\orcid{0000-0001-9408-1101}\inst{\ref{aff28},\ref{aff4}}
\and C.~M.~Gutierrez\orcid{0000-0001-7854-783X}\inst{\ref{aff44},\ref{aff135}}
\and A.~Hall\orcid{0000-0002-3139-8651}\inst{\ref{aff45}}
\and C.~Hern\'andez-Monteagudo\orcid{0000-0001-5471-9166}\inst{\ref{aff135},\ref{aff44}}
\and H.~Hildebrandt\orcid{0000-0002-9814-3338}\inst{\ref{aff136}}
\and J.~Hjorth\orcid{0000-0002-4571-2306}\inst{\ref{aff137}}
\and M.~Jauzac\orcid{0000-0003-1974-8732}\inst{\ref{aff138},\ref{aff81},\ref{aff139},\ref{aff140},\ref{aff91}}
\and S.~Joudaki\orcid{0000-0001-8820-673X}\inst{\ref{aff39}}
\and J.~J.~E.~Kajava\orcid{0000-0002-3010-8333}\inst{\ref{aff141},\ref{aff142},\ref{aff143}}
\and Y.~Kang\orcid{0009-0000-8588-7250}\inst{\ref{aff94}}
\and V.~Kansal\orcid{0000-0002-4008-6078}\inst{\ref{aff144},\ref{aff145}}
\and D.~Karagiannis\orcid{0000-0002-4927-0816}\inst{\ref{aff105},\ref{aff146}}
\and J.~Kim\orcid{0000-0003-2776-2761}\inst{\ref{aff60}}
\and C.~C.~Kirkpatrick\inst{\ref{aff95}}
\and A.~Kov\'acs\orcid{0000-0002-5825-579X}\inst{\ref{aff147},\ref{aff148}}
\and I.~Kova{\v{c}}i{\'{c}}\orcid{0000-0001-6751-3263}\inst{\ref{aff124}}
\and K.~Koyama\orcid{0000-0001-6727-6915}\inst{\ref{aff129}}
\and S.~Kruk\orcid{0000-0001-8010-8879}\inst{\ref{aff26}}
\and M.~C.~Lam\orcid{0000-0002-9347-2298}\inst{\ref{aff45}}
\and F.~Leclercq\orcid{0000-0002-9339-1404}\inst{\ref{aff149}}
\and L.~Legrand\orcid{0000-0003-0610-5252}\inst{\ref{aff150},\ref{aff53}}
\and M.~Lembo\orcid{0000-0002-5271-5070}\inst{\ref{aff149}}
\and G.~Leroy\orcid{0009-0004-2523-4425}\inst{\ref{aff138},\ref{aff81}}
\and G.~F.~Lesci\orcid{0000-0002-4607-2830}\inst{\ref{aff3},\ref{aff4}}
\and J.~Lesgourgues\orcid{0000-0001-7627-353X}\inst{\ref{aff151}}
\and L.~Linke\orcid{0000-0002-2622-8113}\inst{\ref{aff94}}
\and S.~J.~Liu\orcid{0000-0001-7680-2139}\inst{\ref{aff55}}
\and G.~Maggio\orcid{0000-0003-4020-4836}\inst{\ref{aff14}}
\and F.~Mannucci\orcid{0000-0002-4803-2381}\inst{\ref{aff152}}
\and F.~R.~Marleau\orcid{0000-0002-1442-2947}\inst{\ref{aff94}}
\and C.~J.~A.~P.~Martins\orcid{0000-0002-4886-9261}\inst{\ref{aff153},\ref{aff32}}
\and M.~Migliaccio\inst{\ref{aff154},\ref{aff155}}
\and M.~Miluzio\inst{\ref{aff26},\ref{aff156}}
\and G.~Morgante\inst{\ref{aff4}}
\and S.~Nadathur\orcid{0000-0001-9070-3102}\inst{\ref{aff129}}
\and K.~Naidoo\orcid{0000-0002-9182-1802}\inst{\ref{aff129},\ref{aff69}}
\and P.~Natoli\orcid{0000-0003-0126-9100}\inst{\ref{aff105},\ref{aff106}}
\and A.~Navarro-Alsina\orcid{0000-0002-3173-2592}\inst{\ref{aff79}}
\and S.~Nesseris\orcid{0000-0002-0567-0324}\inst{\ref{aff90}}
\and L.~Nicastro\orcid{0000-0001-8534-6788}\inst{\ref{aff4}}
\and F.~Oppizzi\orcid{0000-0003-3904-8370}\inst{\ref{aff5},\ref{aff25},\ref{aff24}}
\and F.~Pace\orcid{0000-0001-8039-0480}\inst{\ref{aff35},\ref{aff36},\ref{aff37},\ref{aff157}}
\and E.~Palazzi\orcid{0000-0002-8691-7666}\inst{\ref{aff4}}
\and D.~Paoletti\orcid{0000-0003-4761-6147}\inst{\ref{aff4},\ref{aff56}}
\and G.~Parimbelli\orcid{0000-0002-2539-2472}\inst{\ref{aff58},\ref{aff27}}
\and F.~Passalacqua\orcid{0000-0002-8606-4093}\inst{\ref{aff25}}
\and K.~Paterson\orcid{0000-0001-8340-3486}\inst{\ref{aff69}}
\and L.~Patrizii\inst{\ref{aff29}}
\and C.~Pattison\orcid{0000-0003-3272-2617}\inst{\ref{aff129}}
\and A.~Pisani\orcid{0000-0002-6146-4437}\inst{\ref{aff54}}
\and D.~Potter\orcid{0000-0002-0757-5195}\inst{\ref{aff158}}
\and G.~W.~Pratt\inst{\ref{aff96}}
\and S.~Quai\orcid{0000-0002-0449-8163}\inst{\ref{aff3},\ref{aff4}}
\and M.~Radovich\orcid{0000-0002-3585-866X}\inst{\ref{aff63}}
\and W.~Roster\orcid{0000-0002-9149-6528}\inst{\ref{aff62}}
\and S.~Sacquegna\orcid{0000-0002-8433-6630}\inst{\ref{aff159}}
\and M.~Sahl\'en\orcid{0000-0003-0973-4804}\inst{\ref{aff160}}
\and D.~B.~Sanders\orcid{0000-0002-1233-9998}\inst{\ref{aff131}}
\and A.~Schneider\orcid{0000-0001-7055-8104}\inst{\ref{aff158}}
\and D.~Sciotti\orcid{0009-0008-4519-2620}\inst{\ref{aff41},\ref{aff80}}
\and E.~Sellentin\orcid{0009-0002-2655-3458}\inst{\ref{aff161},\ref{aff23}}
\and S.~Serjeant\orcid{0000-0002-0517-7943}\inst{\ref{aff162}}
\and L.~C.~Smith\orcid{0000-0002-3259-2771}\inst{\ref{aff163}}
\and I.~Szapudi\orcid{0000-0003-2274-0301}\inst{\ref{aff131}}
\and M.~Talia\orcid{0000-0003-4352-2063}\inst{\ref{aff3},\ref{aff4}}
\and K.~Tanidis\orcid{0000-0001-9843-5130}\inst{\ref{aff164}}
\and C.~Tao\orcid{0000-0001-7961-8177}\inst{\ref{aff54}}
\and F.~Tarsitano\orcid{0000-0002-5919-0238}\inst{\ref{aff165},\ref{aff166},\ref{aff52}}
\and G.~Testera\orcid{0000-0003-2970-766X}\inst{\ref{aff5}}
\and R.~Teyssier\orcid{0000-0001-7689-0933}\inst{\ref{aff167}}
\and S.~Tosi\orcid{0000-0002-7275-9193}\inst{\ref{aff10},\ref{aff1},\ref{aff5}}
\and A.~Troja\orcid{0000-0003-0239-4595}\inst{\ref{aff14}}
\and C.~Uhlemann\orcid{0000-0001-7831-1579}\inst{\ref{aff168},\ref{aff86}}
\and A.~Venhola\orcid{0000-0001-6071-4564}\inst{\ref{aff169}}
\and D.~Vergani\orcid{0000-0003-0898-2216}\inst{\ref{aff4}}
\and S.~Vinciguerra\orcid{0009-0005-4018-3184}\inst{\ref{aff2}}
\and M.~von~Wietersheim-Kramsta\orcid{0000-0003-4986-5091}\inst{\ref{aff81},\ref{aff138}}
\and L.~Wang\orcid{0000-0002-6736-9158}\inst{\ref{aff170},\ref{aff75}}
\and A.~H.~Wright\orcid{0000-0001-7363-7932}\inst{\ref{aff136}}
\and H.~W.~Yeung\orcid{0000-0002-4993-9014}\inst{\ref{aff45}}}
										   
\institute{INAF-Osservatorio Astronomico di Brera, Via Brera 28, 20122 Milano, Italy\label{aff1}
\and
Aix-Marseille Universit\'e, CNRS, CNES, LAM, Marseille, France\label{aff2}
\and
Dipartimento di Fisica e Astronomia "Augusto Righi" - Alma Mater Studiorum Universit\`a di Bologna, via Piero Gobetti 93/2, 40129 Bologna, Italy\label{aff3}
\and
INAF-Osservatorio di Astrofisica e Scienza dello Spazio di Bologna, Via Piero Gobetti 93/3, 40129 Bologna, Italy\label{aff4}
\and
INFN-Sezione di Genova, Via Dodecaneso 33, 16146, Genova, Italy\label{aff5}
\and
Universit\'e de Strasbourg, CNRS, Observatoire astronomique de Strasbourg, UMR 7550, 67000 Strasbourg, France\label{aff6}
\and
Minnesota Institute for Astrophysics, University of Minnesota, 116 Church St SE, Minneapolis, MN 55455, USA\label{aff7}
\and
California Institute of Technology, 1200 E California Blvd, Pasadena, CA 91125, USA\label{aff8}
\and
Jet Propulsion Laboratory, California Institute of Technology, 4800 Oak Grove Drive, Pasadena, CA, 91109, USA\label{aff9}
\and
Dipartimento di Fisica, Universit\`a di Genova, Via Dodecaneso 33, 16146, Genova, Italy\label{aff10}
\and
Ohio University, Physics \& Astronomy Department,1 Ohio University, Athens, OH 45701, USA\label{aff11}
\and
Kavli Institute for the Physics and Mathematics of the Universe (WPI), University of Tokyo, Kashiwa, Chiba 277-8583, Japan\label{aff12}
\and
Dipartimento di Fisica - Sezione di Astronomia, Universit\`a di Trieste, Via Tiepolo 11, 34131 Trieste, Italy\label{aff13}
\and
INAF-Osservatorio Astronomico di Trieste, Via G. B. Tiepolo 11, 34143 Trieste, Italy\label{aff14}
\and
INFN, Sezione di Trieste, Via Valerio 2, 34127 Trieste TS, Italy\label{aff15}
\and
IFPU, Institute for Fundamental Physics of the Universe, via Beirut 2, 34151 Trieste, Italy\label{aff16}
\and
Caltech/IPAC, 1200 E. California Blvd., Pasadena, CA 91125, USA\label{aff17}
\and
Dipartimento di Fisica "Aldo Pontremoli", Universit\`a degli Studi di Milano, Via Celoria 16, 20133 Milano, Italy\label{aff18}
\and
INFN-Sezione di Milano, Via Celoria 16, 20133 Milano, Italy\label{aff19}
\and
Waterloo Centre for Astrophysics, University of Waterloo, Waterloo, Ontario N2L 3G1, Canada\label{aff20}
\and
Department of Physics and Astronomy, University of Waterloo, Waterloo, Ontario N2L 3G1, Canada\label{aff21}
\and
Perimeter Institute for Theoretical Physics, Waterloo, Ontario N2L 2Y5, Canada\label{aff22}
\and
Leiden Observatory, Leiden University, Einsteinweg 55, 2333 CC Leiden, The Netherlands\label{aff23}
\and
Dipartimento di Fisica e Astronomia "G. Galilei", Universit\`a di Padova, Via Marzolo 8, 35131 Padova, Italy\label{aff24}
\and
INFN-Padova, Via Marzolo 8, 35131 Padova, Italy\label{aff25}
\and
ESAC/ESA, Camino Bajo del Castillo, s/n., Urb. Villafranca del Castillo, 28692 Villanueva de la Ca\~nada, Madrid, Spain\label{aff26}
\and
SISSA, International School for Advanced Studies, Via Bonomea 265, 34136 Trieste TS, Italy\label{aff27}
\and
Dipartimento di Fisica e Astronomia, Universit\`a di Bologna, Via Gobetti 93/2, 40129 Bologna, Italy\label{aff28}
\and
INFN-Sezione di Bologna, Viale Berti Pichat 6/2, 40127 Bologna, Italy\label{aff29}
\and
Department of Physics "E. Pancini", University Federico II, Via Cinthia 6, 80126, Napoli, Italy\label{aff30}
\and
INAF-Osservatorio Astronomico di Capodimonte, Via Moiariello 16, 80131 Napoli, Italy\label{aff31}
\and
Instituto de Astrof\'isica e Ci\^encias do Espa\c{c}o, Universidade do Porto, CAUP, Rua das Estrelas, PT4150-762 Porto, Portugal\label{aff32}
\and
Faculdade de Ci\^encias da Universidade do Porto, Rua do Campo de Alegre, 4150-007 Porto, Portugal\label{aff33}
\and
European Southern Observatory, Karl-Schwarzschild-Str.~2, 85748 Garching, Germany\label{aff34}
\and
Dipartimento di Fisica, Universit\`a degli Studi di Torino, Via P. Giuria 1, 10125 Torino, Italy\label{aff35}
\and
INFN-Sezione di Torino, Via P. Giuria 1, 10125 Torino, Italy\label{aff36}
\and
INAF-Osservatorio Astrofisico di Torino, Via Osservatorio 20, 10025 Pino Torinese (TO), Italy\label{aff37}
\and
INAF-IASF Milano, Via Alfonso Corti 12, 20133 Milano, Italy\label{aff38}
\and
Centro de Investigaciones Energ\'eticas, Medioambientales y Tecnol\'ogicas (CIEMAT), Avenida Complutense 40, 28040 Madrid, Spain\label{aff39}
\and
Port d'Informaci\'{o} Cient\'{i}fica, Campus UAB, C. Albareda s/n, 08193 Bellaterra (Barcelona), Spain\label{aff40}
\and
INAF-Osservatorio Astronomico di Roma, Via Frascati 33, 00078 Monteporzio Catone, Italy\label{aff41}
\and
INFN -- Sezione di Napoli, Via Cinthia 6, 80126, Napoli, Italy\label{aff42}
\and
Dipartimento di Fisica e Astronomia "Augusto Righi" - Alma Mater Studiorum Universit\`a di Bologna, Viale Berti Pichat 6/2, 40127 Bologna, Italy\label{aff43}
\and
Instituto de Astrof\'{\i}sica de Canarias, E-38205 La Laguna, Tenerife, Spain\label{aff44}
\and
Institute for Astronomy, University of Edinburgh, Royal Observatory, Blackford Hill, Edinburgh EH9 3HJ, UK\label{aff45}
\and
European Space Agency/ESRIN, Largo Galileo Galilei 1, 00044 Frascati, Roma, Italy\label{aff46}
\and
Universit\'e Claude Bernard Lyon 1, CNRS/IN2P3, IP2I Lyon, UMR 5822, Villeurbanne, F-69100, France\label{aff47}
\and
Institut de Ci\`{e}ncies del Cosmos (ICCUB), Universitat de Barcelona (IEEC-UB), Mart\'{i} i Franqu\`{e}s 1, 08028 Barcelona, Spain\label{aff48}
\and
Instituci\'o Catalana de Recerca i Estudis Avan\c{c}ats (ICREA), Passeig de Llu\'{\i}s Companys 23, 08010 Barcelona, Spain\label{aff49}
\and
Institut de Ciencies de l'Espai (IEEC-CSIC), Campus UAB, Carrer de Can Magrans, s/n Cerdanyola del Vall\'es, 08193 Barcelona, Spain\label{aff50}
\and
UCB Lyon 1, CNRS/IN2P3, IUF, IP2I Lyon, 4 rue Enrico Fermi, 69622 Villeurbanne, France\label{aff51}
\and
Department of Astronomy, University of Geneva, ch. d'Ecogia 16, 1290 Versoix, Switzerland\label{aff52}
\and
Universit\'e Paris-Saclay, CNRS, Institut d'astrophysique spatiale, 91405, Orsay, France\label{aff53}
\and
Aix-Marseille Universit\'e, CNRS/IN2P3, CPPM, Marseille, France\label{aff54}
\and
INAF-Istituto di Astrofisica e Planetologia Spaziali, via del Fosso del Cavaliere, 100, 00100 Roma, Italy\label{aff55}
\and
INFN-Bologna, Via Irnerio 46, 40126 Bologna, Italy\label{aff56}
\and
Institut d'Estudis Espacials de Catalunya (IEEC),  Edifici RDIT, Campus UPC, 08860 Castelldefels, Barcelona, Spain\label{aff57}
\and
Institute of Space Sciences (ICE, CSIC), Campus UAB, Carrer de Can Magrans, s/n, 08193 Barcelona, Spain\label{aff58}
\and
School of Physics, HH Wills Physics Laboratory, University of Bristol, Tyndall Avenue, Bristol, BS8 1TL, UK\label{aff59}
\and
Department of Physics, University of Oxford, Keble Road, Oxford OX1 3RH, UK\label{aff60}
\and
University Observatory, LMU Faculty of Physics, Scheinerstr.~1, 81679 Munich, Germany\label{aff61}
\and
Max Planck Institute for Extraterrestrial Physics, Giessenbachstr. 1, 85748 Garching, Germany\label{aff62}
\and
INAF-Osservatorio Astronomico di Padova, Via dell'Osservatorio 5, 35122 Padova, Italy\label{aff63}
\and
Universit\"ats-Sternwarte M\"unchen, Fakult\"at f\"ur Physik, Ludwig-Maximilians-Universit\"at M\"unchen, Scheinerstr.~1, 81679 M\"unchen, Germany\label{aff64}
\and
Institute of Theoretical Astrophysics, University of Oslo, P.O. Box 1029 Blindern, 0315 Oslo, Norway\label{aff65}
\and
Felix Hormuth Engineering, Goethestr. 17, 69181 Leimen, Germany\label{aff66}
\and
Technical University of Denmark, Elektrovej 327, 2800 Kgs. Lyngby, Denmark\label{aff67}
\and
Cosmic Dawn Center (DAWN), Denmark\label{aff68}
\and
Max-Planck-Institut f\"ur Astronomie, K\"onigstuhl 17, 69117 Heidelberg, Germany\label{aff69}
\and
NASA Goddard Space Flight Center, Greenbelt, MD 20771, USA\label{aff70}
\and
Department of Physics and Astronomy, University College London, Gower Street, London WC1E 6BT, UK\label{aff71}
\and
Universit\'e de Gen\`eve, D\'epartement de Physique Th\'eorique and Centre for Astroparticle Physics, 24 quai Ernest-Ansermet, CH-1211 Gen\`eve 4, Switzerland\label{aff72}
\and
Department of Physics, P.O. Box 64, University of Helsinki, 00014 Helsinki, Finland\label{aff73}
\and
Helsinki Institute of Physics, Gustaf H{\"a}llstr{\"o}min katu 2, University of Helsinki, 00014 Helsinki, Finland\label{aff74}
\and
Kapteyn Astronomical Institute, University of Groningen, PO Box 800, 9700 AV Groningen, The Netherlands\label{aff75}
\and
Laboratoire d'etude de l'Univers et des phenomenes eXtremes, Observatoire de Paris, Universit\'e PSL, Sorbonne Universit\'e, CNRS, 92190 Meudon, France\label{aff76}
\and
SKAO, Jodrell Bank, Lower Withington, Macclesfield SK11 9FT, UK\label{aff77}
\and
Centre de Calcul de l'IN2P3/CNRS, 21 avenue Pierre de Coubertin 69627 Villeurbanne Cedex, France\label{aff78}
\and
Universit\"at Bonn, Argelander-Institut f\"ur Astronomie, Auf dem H\"ugel 71, 53121 Bonn, Germany\label{aff79}
\and
INFN-Sezione di Roma, Piazzale Aldo Moro, 2 - c/o Dipartimento di Fisica, Edificio G. Marconi, 00185 Roma, Italy\label{aff80}
\and
Department of Physics, Institute for Computational Cosmology, Durham University, South Road, Durham, DH1 3LE, UK\label{aff81}
\and
Institute of Physics, Laboratory of Astrophysics, Ecole Polytechnique F\'ed\'erale de Lausanne (EPFL), Observatoire de Sauverny, 1290 Versoix, Switzerland\label{aff82}
\and
Telespazio UK S.L. for European Space Agency (ESA), Camino bajo del Castillo, s/n, Urbanizacion Villafranca del Castillo, Villanueva de la Ca\~nada, 28692 Madrid, Spain\label{aff83}
\and
Institut de F\'{i}sica d'Altes Energies (IFAE), The Barcelona Institute of Science and Technology, Campus UAB, 08193 Bellaterra (Barcelona), Spain\label{aff84}
\and
European Space Agency/ESTEC, Keplerlaan 1, 2201 AZ Noordwijk, The Netherlands\label{aff85}
\and
School of Mathematics, Statistics and Physics, Newcastle University, Herschel Building, Newcastle-upon-Tyne, NE1 7RU, UK\label{aff86}
\and
Space Science Data Center, Italian Space Agency, via del Politecnico snc, 00133 Roma, Italy\label{aff87}
\and
Centre National d'Etudes Spatiales -- Centre spatial de Toulouse, 18 avenue Edouard Belin, 31401 Toulouse Cedex 9, France\label{aff88}
\and
Institute of Space Science, Str. Atomistilor, nr. 409 M\u{a}gurele, Ilfov, 077125, Romania\label{aff89}
\and
Instituto de F\'isica Te\'orica UAM-CSIC, Campus de Cantoblanco, 28049 Madrid, Spain\label{aff90}
\and
Institut de Recherche en Astrophysique et Plan\'etologie (IRAP), Universit\'e de Toulouse, CNRS, UPS, CNES, 14 Av. Edouard Belin, 31400 Toulouse, France\label{aff91}
\and
Universit\'e St Joseph; Faculty of Sciences, Beirut, Lebanon\label{aff92}
\and
Departamento de F\'isica, FCFM, Universidad de Chile, Blanco Encalada 2008, Santiago, Chile\label{aff93}
\and
Universit\"at Innsbruck, Institut f\"ur Astro- und Teilchenphysik, Technikerstr. 25/8, 6020 Innsbruck, Austria\label{aff94}
\and
Department of Physics and Helsinki Institute of Physics, Gustaf H\"allstr\"omin katu 2, University of Helsinki, 00014 Helsinki, Finland\label{aff95}
\and
Universit\'e Paris-Saclay, Universit\'e Paris Cit\'e, CEA, CNRS, AIM, 91191, Gif-sur-Yvette, France\label{aff96}
\and
Departamento de F\'isica, Faculdade de Ci\^encias, Universidade de Lisboa, Edif\'icio C8, Campo Grande, PT1749-016 Lisboa, Portugal\label{aff97}
\and
Instituto de Astrof\'isica e Ci\^encias do Espa\c{c}o, Faculdade de Ci\^encias, Universidade de Lisboa, Tapada da Ajuda, 1349-018 Lisboa, Portugal\label{aff98}
\and
Mullard Space Science Laboratory, University College London, Holmbury St Mary, Dorking, Surrey RH5 6NT, UK\label{aff99}
\and
Cosmic Dawn Center (DAWN)\label{aff100}
\and
Niels Bohr Institute, University of Copenhagen, Jagtvej 128, 2200 Copenhagen, Denmark\label{aff101}
\and
Universidad Polit\'ecnica de Cartagena, Departamento de Electr\'onica y Tecnolog\'ia de Computadoras,  Plaza del Hospital 1, 30202 Cartagena, Spain\label{aff102}
\and
European University of Technology EUt+, European Union\label{aff103}
\and
Astronomisches Rechen-Institut, Zentrum f\"ur Astronomie der Universit\"at Heidelberg, M\"onchhofstr. 12-14, 69120 Heidelberg, Germany\label{aff104}
\and
Dipartimento di Fisica e Scienze della Terra, Universit\`a degli Studi di Ferrara, Via Giuseppe Saragat 1, 44122 Ferrara, Italy\label{aff105}
\and
Istituto Nazionale di Fisica Nucleare, Sezione di Ferrara, Via Giuseppe Saragat 1, 44122 Ferrara, Italy\label{aff106}
\and
Universit\'e Paris Cit\'e, CNRS, Astroparticule et Cosmologie, 75013 Paris, France\label{aff107}
\and
INAF, Istituto di Radioastronomia, Via Piero Gobetti 101, 40129 Bologna, Italy\label{aff108}
\and
Astronomical Observatory of the Autonomous Region of the Aosta Valley (OAVdA), Loc. Lignan 39, I-11020, Nus (Aosta Valley), Italy\label{aff109}
\and
Universit\'e C\^{o}te d'Azur, Observatoire de la C\^{o}te d'Azur, CNRS, Laboratoire Lagrange, Bd de l'Observatoire, CS 34229, 06304 Nice cedex 4, France\label{aff110}
\and
Department of Mathematical Physics, Institute of Physics, University of S\~ao Paulo, R. do Mat\~ao 1371, 05508-090, S\~ao Paulo, SP, Brazil\label{aff111}
\and
CEA Saclay, DFR/IRFU, Service d'Astrophysique, Bat. 709, 91191 Gif-sur-Yvette, France\label{aff112}
\and
Univ. Grenoble Alpes, CNRS, Grenoble INP, LPSC-IN2P3, 53, Avenue des Martyrs, 38000 Grenoble, France\label{aff113}
\and
Dipartimento di Fisica, Sapienza Universit\`a di Roma, Piazzale Aldo Moro 2, 00185 Roma, Italy\label{aff114}
\and
Department of Mathematics and Physics E. De Giorgi, University of Salento, Via per Arnesano, CP-I93, 73100, Lecce, Italy\label{aff115}
\and
INFN, Sezione di Lecce, Via per Arnesano, CP-193, 73100, Lecce, Italy\label{aff116}
\and
INAF-Sezione di Lecce, c/o Dipartimento Matematica e Fisica, Via per Arnesano, 73100, Lecce, Italy\label{aff117}
\and
Institut d'Astrophysique de Paris, 98bis Boulevard Arago, 75014, Paris, France\label{aff118}
\and
ICL, Junia, Universit\'e Catholique de Lille, LITL, 59000 Lille, France\label{aff119}
\and
ICSC - Centro Nazionale di Ricerca in High Performance Computing, Big Data e Quantum Computing, Via Magnanelli 2, Bologna, Italy\label{aff120}
\and
CERCA/ISO, Department of Physics, Case Western Reserve University, 10900 Euclid Avenue, Cleveland, OH 44106, USA\label{aff121}
\and
Laboratoire Univers et Th\'eorie, Observatoire de Paris, Universit\'e PSL, Universit\'e Paris Cit\'e, CNRS, 92190 Meudon, France\label{aff122}
\and
Departamento de F{\'\i}sica Fundamental. Universidad de Salamanca. Plaza de la Merced s/n. 37008 Salamanca, Spain\label{aff123}
\and
Universiteit Gent, Department of Physics and Astronomy, Proeftuinstraat 86 N3, 9000 Ghent, Belgium
\label{aff124}
\and
Aix-Marseille Universit\'e, Universit\'e de Toulon, CNRS, CPT, Marseille, France\label{aff125}
\and
Center for Data-Driven Discovery, Kavli IPMU (WPI), UTIAS, The University of Tokyo, Kashiwa, Chiba 277-8583, Japan\label{aff126}
\and
Jodrell Bank Centre for Astrophysics, Department of Physics and Astronomy, University of Manchester, Oxford Road, Manchester M13 9PL, UK\label{aff127}
\and
Department of Physics, Astronomy and Mathematics, University of Hertfordshire, College Lane, Hatfield AL10 9AB, UK\label{aff128}
\and
Institute of Cosmology and Gravitation, University of Portsmouth, Portsmouth PO1 3FX, UK\label{aff129}
\and
Department of Physics \& Astronomy, University of California Irvine, Irvine CA 92697, USA\label{aff130}
\and
Institute for Astronomy, University of Hawaii, 2680 Woodlawn Drive, Honolulu, HI 96822, USA\label{aff131}
\and
Departamento F\'isica Aplicada, Universidad Polit\'ecnica de Cartagena, Campus Muralla del Mar, 30202 Cartagena, Murcia, Spain\label{aff132}
\and
Instituto de F\'isica de Cantabria, Edificio Juan Jord\'a, Avenida de los Castros, 39005 Santander, Spain\label{aff133}
\and
Department of Computer Science, Aalto University, PO Box 15400, Espoo, FI-00 076, Finland\label{aff134}
\and
Universidad de La Laguna, Dpto. Astrof\'\i sica, E-38206 La Laguna, Tenerife, Spain\label{aff135}
\and
Ruhr University Bochum, Faculty of Physics and Astronomy, Astronomical Institute (AIRUB), German Centre for Cosmological Lensing (GCCL), 44780 Bochum, Germany\label{aff136}
\and
DARK, Niels Bohr Institute, University of Copenhagen, Jagtvej 155, 2200 Copenhagen, Denmark\label{aff137}
\and
Department of Physics, Centre for Extragalactic Astronomy, Durham University, South Road, Durham, DH1 3LE, UK\label{aff138}
\and
Astrophysics Research Centre, University of KwaZulu-Natal, Westville Campus, Durban 4041, South Africa\label{aff139}
\and
School of Mathematics, Statistics \& Computer Science, University of KwaZulu-Natal, Westville Campus, Durban 4041, South Africa\label{aff140}
\and
Department of Physics and Astronomy, Vesilinnantie 5, University of Turku, 20014 Turku, Finland\label{aff141}
\and
Finnish Centre for Astronomy with ESO (FINCA), Quantum, Vesilinnantie 5, University of Turku, 20014 Turku, Finland\label{aff142}
\and
Serco for European Space Agency (ESA), Camino bajo del Castillo, s/n, Urbanizacion Villafranca del Castillo, Villanueva de la Ca\~nada, 28692 Madrid, Spain\label{aff143}
\and
ARC Centre of Excellence for Dark Matter Particle Physics, Melbourne, Australia\label{aff144}
\and
Centre for Astrophysics \& Supercomputing, Swinburne University of Technology,  Hawthorn, Victoria 3122, Australia\label{aff145}
\and
Department of Physics and Astronomy, University of the Western Cape, Bellville, Cape Town, 7535, South Africa\label{aff146}
\and
MTA-CSFK Lend\"ulet Large-Scale Structure Research Group, Konkoly-Thege Mikl\'os \'ut 15-17, H-1121 Budapest, Hungary\label{aff147}
\and
Konkoly Observatory, HUN-REN CSFK, MTA Centre of Excellence, Budapest, Konkoly Thege Mikl\'os {\'u}t 15-17. H-1121, Hungary\label{aff148}
\and
Institut d'Astrophysique de Paris, UMR 7095, CNRS, and Sorbonne Universit\'e, 98 bis boulevard Arago, 75014 Paris, France\label{aff149}
\and
Brazilian Center for Research in Physics (CBPF), Dr. Xavier Sigaud st. 150, zip 22290-180, Rio de Janeiro, RJ, Brazil\label{aff150}
\and
Institute for Theoretical Particle Physics and Cosmology (TTK), RWTH Aachen University, 52056 Aachen, Germany\label{aff151}
\and
INAF-Osservatorio Astrofisico di Arcetri, Largo E. Fermi 5, 50125, Firenze, Italy\label{aff152}
\and
Centro de Astrof\'{\i}sica da Universidade do Porto, Rua das Estrelas, 4150-762 Porto, Portugal\label{aff153}
\and
Dipartimento di Fisica, Universit\`a di Roma Tor Vergata, Via della Ricerca Scientifica 1, Roma, Italy\label{aff154}
\and
INFN, Sezione di Roma 2, Via della Ricerca Scientifica 1, Roma, Italy\label{aff155}
\and
HE Space for European Space Agency (ESA), Camino bajo del Castillo, s/n, Urbanizacion Villafranca del Castillo, Villanueva de la Ca\~nada, 28692 Madrid, Spain\label{aff156}
\and
Instituto de Astrof\'isica e Ci\^encias do Espa\c{c}o, Faculdade de Ci\^encias, Universidade de Lisboa, Campo Grande, 1749-016 Lisboa, Portugal\label{aff157}
\and
Department of Astrophysics, University of Zurich, Winterthurerstrasse 190, 8057 Zurich, Switzerland\label{aff158}
\and
INAF - Osservatorio Astronomico d'Abruzzo, Via Maggini, 64100, Teramo, Italy\label{aff159}
\and
Theoretical astrophysics, Department of Physics and Astronomy, Uppsala University, Box 516, 751 37 Uppsala, Sweden\label{aff160}
\and
Mathematical Institute, University of Leiden, Einsteinweg 55, 2333 CA Leiden, The Netherlands\label{aff161}
\and
School of Physical Sciences, The Open University, Milton Keynes, MK7 6AA, UK\label{aff162}
\and
Institute of Astronomy, University of Cambridge, Madingley Road, Cambridge CB3 0HA, UK\label{aff163}
\and
Center for Astrophysics and Cosmology, University of Nova Gorica, Nova Gorica, Slovenia\label{aff164}
\and
Kobayashi-Maskawa Institute for the Origin of Particles and the Universe, Nagoya University, Chikusa-ku, Nagoya, 464-8602, Japan\label{aff165}
\and
Institute for Particle Physics and Astrophysics, Dept. of Physics, ETH Zurich, Wolfgang-Pauli-Strasse 27, 8093 Zurich, Switzerland\label{aff166}
\and
Department of Astrophysical Sciences, Peyton Hall, Princeton University, Princeton, NJ 08544, USA\label{aff167}
\and
Fakult\"at f\"ur Physik, Universit\"at Bielefeld, Postfach 100131, 33501 Bielefeld, Germany\label{aff168}
\and
Space physics and astronomy research unit, University of Oulu, Pentti Kaiteran katu 1, FI-90014 Oulu, Finland\label{aff169}
\and
SRON Netherlands Institute for Space Research, Landleven 12, 9747 AD, Groningen, The Netherlands\label{aff170}}

\title{\Euclid}
\subtitle{Bypass end-to-end simulations for spectroscopic galaxy clustering performance assessment}    

\titlerunning{Bypass spectroscopic simulations}
\authorrunning{Euclid Collaboration: B.~R.~Granett et al.}




\abstract{
We present a simulation framework for \Euclid galaxy clustering measurements that reproduces the essential behaviour of the spectroscopic reduction chain, bypassing the computation required by full pixel-level simulations. The framework comprises two complementary tools, \fastspec, which simulates 1-dimensional slitless spectra with a simplified instrument and noise model, and \pypelid, which computes detection probabilities on input catalogues, accounting for the survey dithering sequence and spatially varying noise. We develop calibrated prescriptions for the survey selection function, redshift errors, and redshift interloper populations. Applied to the Euclid Wide Survey, the baseline selection yields a redshift purity of 74\% and a sample completeness of 45\%, which improves with the use of \Euclid photometric colours. The redshift error distribution is found to have a Gaussian core of width $\sigma_z = 9.4 \times 10^{-4}$, with no significant trend over $0.9 < z < 1.8$. Emission lines [\ion{O}{iii}]$\lambda$5008 and [\ion{S}{iii}]$\lambda$9530 misidentified as \Ha contribute a few to ten per cent depending on redshift, while catastrophic redshift measurement failures dominate at low signal-to-noise ratio.  Simulating the Data Release 1 footprint, we find that complex survey selection effects can be corrected with a forward-modelled random catalogue, recovering the monopole and quadrupole of the galaxy 2-point correlation function. The simplified noise model does not include residual contamination from overlapping spectra, or other artefacts seen in early \Euclid data; this will be revisited with DR1 data. The framework provides an efficient route for characterising the \Euclid spectroscopic selection function at the precision required for galaxy clustering science, and is the basis for systematic error assessments in \Euclid galaxy clustering analyses.
}


\maketitle

\nolinenumbers
%
%

\section{Introduction}

The European Space Agency's \Euclid mission is currently performing an imaging and spectroscopic redshift survey designed to probe the late-time acceleration of the expansion of the Universe \citep{EuclidSkyOverview}. In preparation for forthcoming \Euclid science analyses, this paper presents simulation methods tailored for the spectroscopic galaxy survey to enable the measurement of 3-dimensional galaxy clustering statistics. Two primary physical phenomena will be inferred from \Euclid's 2-point clustering measurements: the expansion history of the Universe via baryon acoustic oscillations (BAOs); and the growth rate of cosmic structure via redshift-space distortions (RSDs). Additionally, these 3-dimensional clustering measurements will be combined with \Euclid's weak lensing and photometric clustering probes. However, reaching the precision promised by the \Euclid mission demands exquisite control over systematic uncertainties that could otherwise corrupt the cosmological signal \citep{EP-Monaco2}. Such systematic uncertainties arise at every stage of the analysis, spanning instrumentation and calibration effects, astrophysical foregrounds, and data-processing pipelines. Beyond observational systematic errors, uncertainties in theoretical modelling and observable covariances can introduce additional biases. Given the inherent complexity of these analyses, one way to validate them is to process simulated data through the data reduction and analysis pipeline. These so-called end-to-end simulations have become an indispensable tool to quantify the impact of known systematic errors and validate the robustness of the final cosmological inference.

\Euclid's survey strategy introduces a variety of selection effects into the spectroscopic sample, which propagate as systematic errors in galaxy clustering statistics. Notably, \Euclid is the first wide-area galaxy redshift survey to utilise a space-based, slitless spectrograph design. Slitless spectroscopy captures the dispersed light from every source within the field of view without requiring an intervening slit mask or optical fibres for target selection. While this technique is exceptionally efficient for emission-line galaxy surveys \citep{Atek2010, Momcheva2016}, it comes with complications. The spectra of nearby sources overlap on the detector, which can lead to contamination of the signal in the spectral extraction process.  In addition to the 1$^{\rm st}$-order spectra, the grism dispersive element generates compact 0$^{\rm th}$-order images that can masquerade as emission lines. The superposition of spectra on the detector degrades both spectral extraction and redshift determination \citep{Q1-TP006, Q1-TP007}. Spectral contamination can obscure genuine features or introduce spurious emission line and continuum signals, ultimately leading to redshift measurement failures. The impact of systematic redshift errors on \Euclid clustering measurements was investigated in configuration and Fourier space (\citeauthor{EP-Risso} 2026; Euclid Collaboration: Lee et al. in prep). Furthermore, the survey selection function exhibits angular inhomogeneities across the sky due to varying noise levels and spectro-photometric calibration uncertainties \citep{Colavincenzo2017, Markovic2017}. If left unmitigated, these spatial variations can bias the cosmological parameters inferred from the 3-dimensional galaxy clustering signal. Modelling these effects analytically is challenging, since they depend on an intricate interplay between instrumental response, astrophysical foregrounds, and the intrinsic properties of the underlying galaxy distribution. Consequently, a comprehensive assessment requires end-to-end simulations that encompass both the survey strategy and the subsequent data analysis pipeline \citep{EP-Monaco2}.

The \Euclid ground segment includes a dedicated organisational unit (OU) for producing survey simulations \citep[OU-SIM,][]{EP-Serrano}, which has developed comprehensive simulators for both imaging and spectroscopic data products in parallel with the data reduction pipeline. These simulations have long been integral to the mission for performance verification and integration testing. Recently, a major simulation campaign was conducted, as described in \citet{EP-Passalacqua}, which generated synthetic spectroscopic data products and processed them through the complete, end-to-end data reduction pipeline. This represents a significant advancement over previous campaigns, which only applied the spectral extraction step to simulated data \citep{Gabarra-EP31,EP-Lusso}. Computational resources are not available to simulate and reduce multiple realisations of survey areas of thousands of square degrees needed for galaxy clustering analyses. This computational hurdle led us to develop bypass simulation and reduction pipelines that use proxies in place of particular steps, with the goal of preserving the essential characteristics of the survey selection function, but with reduced computational cost. 

This paper describes the various tools developed as part of this end-to-end simulation framework. 
\fastspec is designed to simulate slitless spectroscopic observations of individual galaxies. To assess selection effects across the survey, we developed \pypelid, which processes mock galaxy catalogues and directly outputs detection probabilities and simulated redshift measurements. In addition to these general-purpose simulation tools, we developed analytical prescriptions to efficiently imprint the survey selection function directly onto mock galaxy catalogues, while treating the angular and radial selection effects independently. The angular selection effects were characterised using a tabulated detection probability mapping a reduced set of observational parameters to a final detection probability. This approach was employed to imprint the selection function onto \num{1000} realisations of the EuclidLargeMocks catalogue from \citet{EP-Monaco1}. For the radial selection function and associated redshift uncertainties, we developed a model to characterise the redshift error probability distribution function. This model, presented in this paper, was applied to assign realistic redshift errors to the mock catalogues used in \citet{EP-Risso} and \citelee.

The paper is organised as follows. Section~2 outlines the \Euclid spectroscopic pipeline. Section 3 presents our end-to-end simulation framework, including the input catalogues, instrument model, and simulation codes. Section~4 presents the resulting reference survey performance in terms of redshift purity and sample completeness from bypass simulations. Section~5 presents the detection model, redshift errors, and redshift interloper prescriptions. Section~6 presents a realistic survey simulation constructed with \pypelid and Section~7 discusses our results.

\section{Euclid spectroscopic pipeline}
\begin{figure}
    \vspace{-10pt}
    \centering
    \includegraphics[width=3.5in]{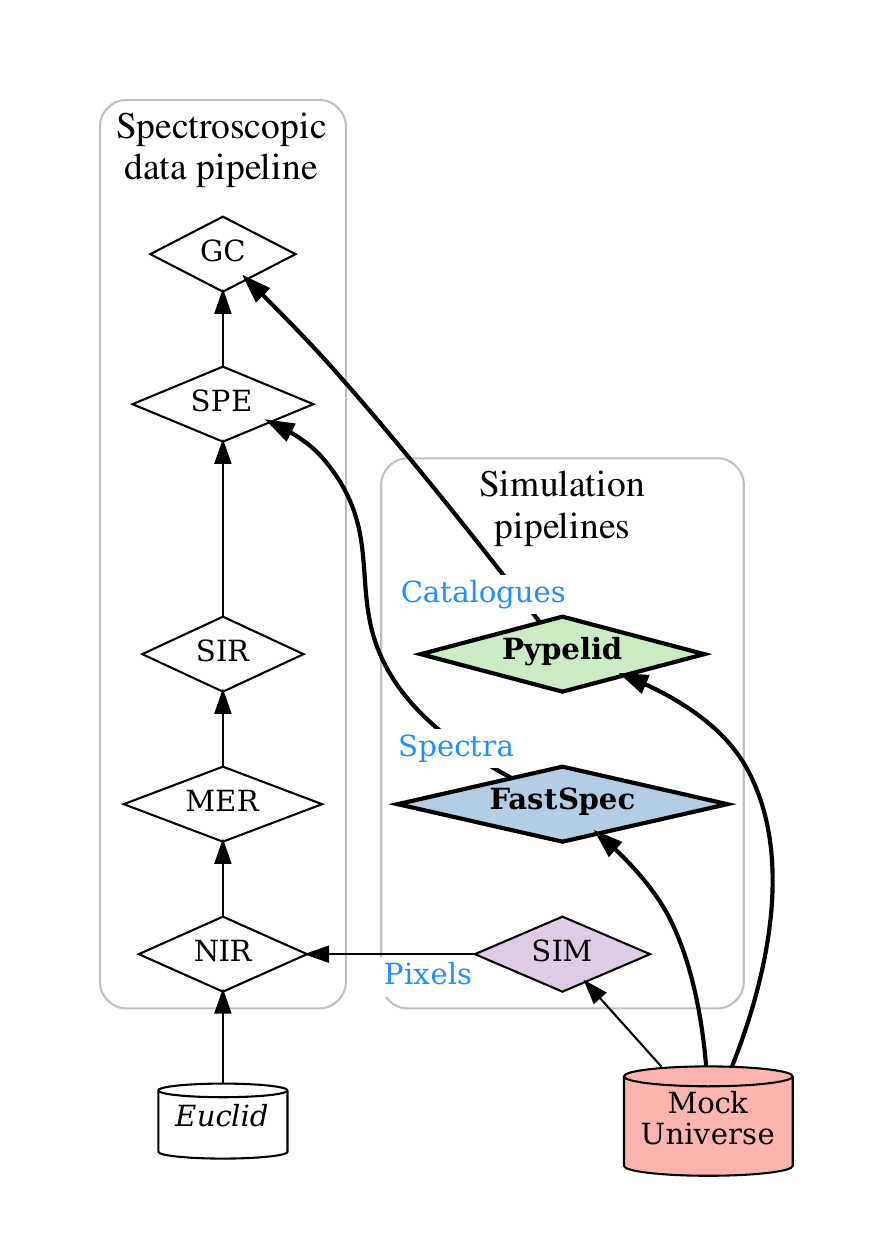}
    \vspace{-15pt}
    \caption{Data flow through the spectroscopic reduction pipeline from \Euclid through the  processing chain: NIR, MER, SIR, SPE, and GC. Three simulation pipelines, OU-SIM, \fastspec, and \pypelid, inject simulated data products at different steps.
    \fastspec and \pypelid bypass the most computationally expensive steps of simulating and processing the pixel-level imaging and spectroscopic exposures.}
    \label{fig:bypassflow}
\end{figure}

The \Euclid data reduction pipeline processes the imaging and spectroscopic data from the satellite, producing high-level products that include the source catalogues, as well as galaxy clustering summary statistics for cosmological analyses.
The pipeline is illustrated in Fig.~\ref{fig:bypassflow}. The significant flow of data (roughly 20 GB per hour) from \Euclid's near-infrared spectrometer and photometer \citep[NISP,][]{EuclidSkyNISP} and visible camera \citep[VIS,][]{EuclidSkyVIS} is processed by the Science Ground Segment (SGS) in specific OUs. The raw photometric and spectroscopic exposures are processed by OU-NIR to apply calibrations including dark current and flat-fielding corrections \citep{Q1-TP003}. After initial calibrations by OU-NIR and OU-VIS \citep{Q1-TP002}, individual imaging frames are combined by OU-MER for source detection and measurement \citep{Q1-TP004}. The resulting catalogue provides the input for spectral extraction. OU-SIR extracts the 2-dimensional spectra and corrects for contamination from overlapping sources \citep{Q1-TP006}. These cleaned spectra are reduced to 1-dimensional traces, flux calibrated, and eventually co-added into a single 1-dimensional spectrum for each object. OU-SPE then estimates the redshift and measures spectral features using a template-fitting approach \citep{Q1-TP007}. Finally, the compiled redshift catalogue is delivered to the higher-level sample selection, random catalogue, and galaxy clustering processing functions in OU-LE3 \citep[e.g.][]{EP-delaTorre}. In parallel, simulations of the raw data are produced by OU-SIM \citep{EP-Serrano}.

The \fastspec and \pypelid codes bypass steps in the data processing chain as indicated in Fig.~\ref{fig:bypassflow}. \fastspec simulates the calibrated 2-dimensional and 1-dimensional extracted spectroscopic data at the intermediate stage of the pipeline, which can then be processed by OU-SPE. Crucially, it bypasses the full spectral-extraction processing function within the standard data-reduction pipeline, enabling rapid testing. \pypelid, instead, bypasses the full spectroscopic data reduction pipeline, and from an input mock galaxy catalogue, outputs a catalogue containing detection probabilities and redshift measurement errors ready for galaxy clustering analysis. The detection model at the heart of \pypelid needs to be calibrated on a set of reference simulated spectra processed through the OU-SPE redshift measurement pipeline.

\section{Bypass simulation pipelines}

\subsection{Input catalogues}
 \begin{figure}
    \includegraphics[width=3.5in]{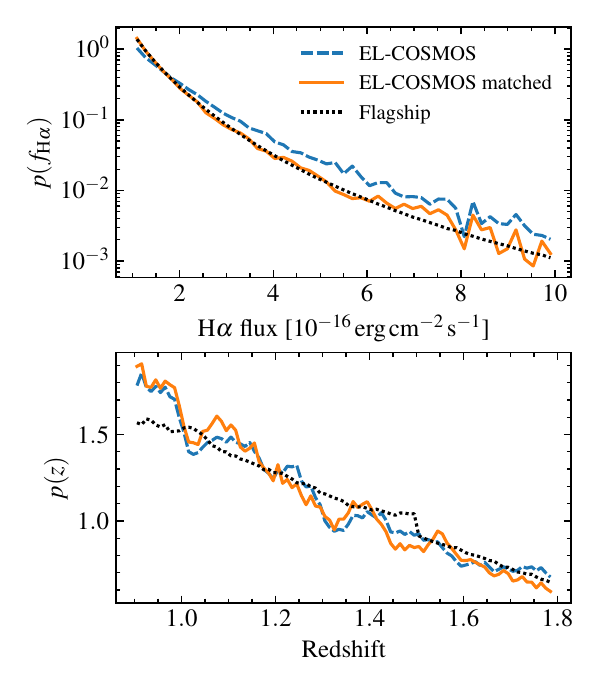}
        \caption{Flux (top) and redshift (bottom) distributions of the EL-COSMOS and Flagship galaxy catalogues (dashed and dotted lines). We applied weights to the EL-COSMOS catalogue to match the \citet{Pozzetti2016} `model 3' \Ha luminosity function (solid line), which was adopted for the Flagship mock catalogue. \label{fig:nz}}
\end{figure}

\subsubsection{EL-COSMOS catalogue}
\label{sec:elcosmos}

Realistic galaxy catalogues with spectro-photometric properties are needed as input to simulate the survey. For this, we used an updated version of the EL-COSMOS catalogue from \citet{Saito2020}, obtained with the same method but using as input the COSMOS-2020 photometric catalogue \citep{Weaver2022}, which supersedes the COSMOS-2015 catalogue \citep{Laigle2016} that was originally used. The photometric and spectroscopic information for each galaxy was modelled with synthetic spectral energy distributions that include both nebular and stellar contributions. The nebular line fluxes were modelled using the method of \citet{Schaerer2009}, accounting for the fact that the amount of dust attenuation for emission lines from nebular regions is different from that for the stellar continuum. Overall, this allowed the production of realistic spectra and emission-line fluxes associated with all COSMOS-2020 galaxies. The EL-COSMOS catalogue comprises in total \num{205314} objects with associated emission-line fluxes and synthetic spectra, as well as all photometrically derived galaxy physical properties from \citet{Weaver2022}. The \Ha luminosity function in the EL-COSMOS catalogue estimated in \citet{Saito2020} differs from the \citet{Pozzetti2016} models used for \Euclid forecasts \citep{Blanchard-EP7}. We therefore applied weights to EL-COSMOS galaxies to match flux distributions, as shown in Fig.~\ref{fig:nz}.

\subsubsection{Flagship mock galaxy catalogue}\label{sec:flagship}

In addition to the EL-COSMOS catalogue, we used the Euclid Flagship mock galaxy catalogue. The latter is based on the Flagship $N$-body simulation \citep{pkdgrav3} that has a particle mass resolution of $m_{\rm p} = 10^9\,h^{-1}\,\si{\solarmass}$ and a volume of $(3.6\,h^{-1}\,\mathrm{Gpc})^3$. 
Galaxies were added to the $N$-body simulation using a halo occupation distribution (HOD) approach and physical properties simulated using empirical relations calibrated on observational data. Details on the methods employed to generate the mock galaxy catalogue can be found in \cite{EuclidSkyFlagship}. Although less realistic in terms of galaxy physical properties than the EL-COSMOS catalogue, the Flagship mock galaxy catalogue covers an octant of the sky and thus allows us to simulate much larger volumes. The Flagship mock galaxy catalogue was accessed from the CosmoHub platform\footnote{https://cosmohub.pic.es} \citep{Carretero2017, Tallada2020}.

\subsection{Reference Euclid Wide and Deep Surveys}

\begin{figure*}
    \includegraphics[width=2.in]{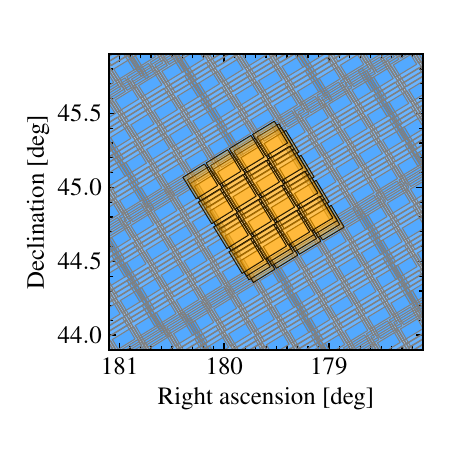}
    \includegraphics[width=5.3in]{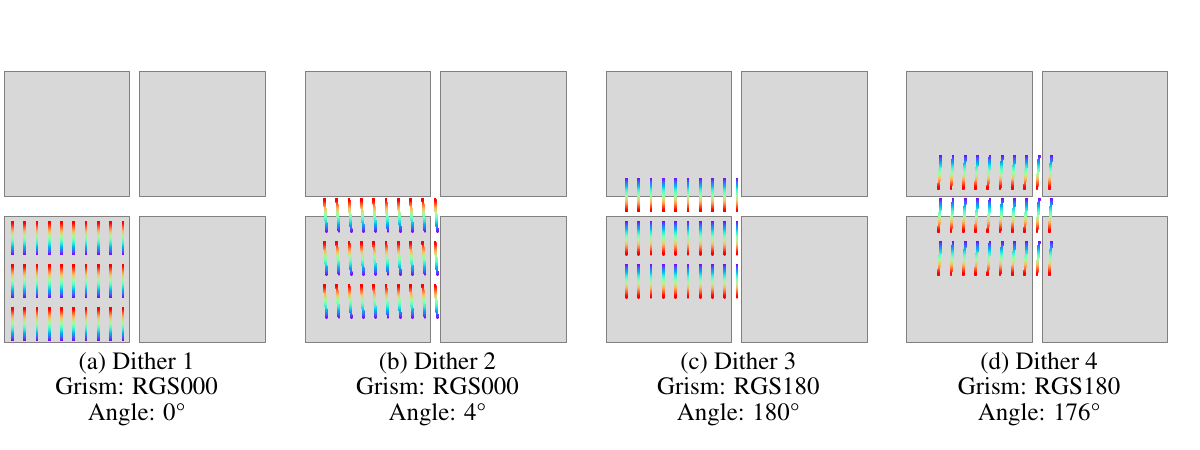}
        \caption{\emph{Left panel}: illustration of an area of the EWS. The polygons represent the outline of detectors of the NISP focal plane array projected on the sky. The focal plane is aligned with the ecliptic pole. Highlighted in orange is the footprint of one NISP-S dither sequence consisting of four exposures. 
        \emph{Right 4 panels}: zoom-in on one quadrant of the NISP focal plane. The grey rectangles represent 4 detectors. The line segments mark the spectral traces of sources arranged in a fixed grid on the sky. The rainbow colour gradient, purple to red, indicates wavelength from 1.2--1.9\,\micron.  Panels (a)--(d) show the position and orientation of the spectral traces at each step of the dither sequence with the RGS000 and RGS180 grisms in 0\degr and $\pm 4\degr$ tilted configurations. \label{fig:surveypattern}}
\end{figure*}

The Euclid Wide Survey (EWS)  will cover approximately \num{14000}~deg$^2$ of the sky \citep{Scaramella-EP1}. The survey strategy uses a four-dither exposure pattern to fill the gaps between detectors \citep[see section~4 of][]{Scaramella-EP1}, referred to as the reference observing sequence (ROS). Figure~\ref{fig:surveypattern} illustrates the reference pattern of NISP spectroscopic (NISP-S hereafter) exposures and the orientation of the spectra in each exposure, according to the grism configuration. In a ROS, each NISP-S exposure in the dithering sequence is made with the red grism RGS000 ($0\degr$) or RGS180 ($180\degr$), tilted by a $4\degr$ rotation of the grism wheel. The sequence follows RGS000+$0\degr$, RGS000+$4\degr$, RGS180+$0\degr$, RGS180-$4\degr$. The Euclid Deep Survey \citep[EDS,][]{EuclidSkyOverview} also uses blue grism observations and its strategy consists of at least $15$ red and $25$ blue grism visits over the same areas, yielding a depth approximately 6 times fainter in flux than in the EWS. The visits are made throughout the year at different position angles such that the dispersion direction in the exposures varies.

\subsection{Instrument model}

A number of instrumental effects must be taken into account to properly simulate a spectroscopic exposure of a galaxy in \Euclid.
These are captured in the instrumental response function, wavelength dispersion, and point-spread function (PSF).
We model the effective instrumental response function $R(\lambda)$ including instrumental efficiencies and Galactic extinction as
\begin{equation}\label{eq:response}
R(\lambda) = A_{\rm col}\,t_{\rm exp}\,q_{\rm e}(\lambda)\ 10^{-0.4\,A_{\rm MW}(\lambda)}\, T(\lambda)\, \frac{\lambda}{\hP \, c}\, ,
\end{equation}
where $\lambda$ is the wavelength, $A_{\rm col}$ is the telescope collecting area, $t_{\rm exp}$ is the exposure time, $A_{\rm MW}$ is the Milky Way extinction curve, $q_{\rm e}$ is the quantum efficiency, $T$ is the total spectrograph transmission function, $\hP$ is the Planck constant, and $c$ is the speed of light in vacuum. The integrated signal from an astrophysical source with spectral energy distribution $f_{\lambda}(\lambda)$ recorded by a detector is given in count units by integrating over the transmission window $\lambda_{\rm start}<\lambda<\lambda_{\rm end}$ as
\begin{equation}\label{eq:signal}
s = \int_{\lambda_{\rm start}}^{\lambda_{\rm end}} \,R(\lambda)\,f_{\lambda}(\lambda)\,\diff \lambda\,.
\end{equation}
The spectrograph maps the wavelength $\lambda$ to a position on the detector $\vec{x}(\lambda)$. This mapping can be approximated as linear, such that
\begin{equation}
    \lambda = d_\lambda \left| \vec{x} - \vec{x}_0 \right| + \lambda_0 \, ,
\end{equation}
where $d_\lambda = \Delta \lambda/\Delta x$ is the wavelength dispersion, $\vec{x}_0$ is one end of the spectral trace on the detector, and $\lambda_0$ is a constant.

The dispersed image recorded on the detector $s_{2 {\rm D}}(\vec{x})$ is obtained by convolving the 1-dimensional spectral trace on the focal plane, $s(\vec{x})$, with the galaxy image, $g(\vec{x})$, and in turn with the PSF, $o(\vec{x})$, where $\vec{x}=(x,y)$ are detector pixel coordinates, as 
\begin{equation}\label{eq:convolution}
    s_{2 {\rm D}} = s * g * o \, ,
\end{equation}
where $*$ denotes the convolution product. The adopted model PSF is radially symmetric and described by a two-component Gaussian mixture model given in Appendix~\ref{app:psf}.

The 1-dimensional spectrum that is used for the spectroscopic redshift measurement is obtained by a weighted sum of pixels in the dispersed image. Considering the pixel indices $i$ aligned with the dispersion direction and $j$ perpendicular to it, the extraction is performed by summing over $n$ pixels in the perpendicular direction with a normalised weight function $w_j$,
\begin{equation} \label{eq:1dextraction}
    s_{1{\rm D},\, i} = \sum_{j=1}^{n} \, w_j \, s_{2{\rm D},\, ij}\, .
\end{equation}
To reproduce the extraction procedure implemented prior to launch \citep{Q1-TP006} we used an aperture with constant weights for extraction.\footnote{The OU-SIR spectroscopic extraction pipeline was subsequently updated to use extraction weights based on the source profile to optimise the signal-to-noise ratio.} After extraction, multiple exposures may be combined with inverse-variance weighting. The details of the implementation in \pypelid are provided in Appendix~\ref{app:pypelid}.

\subsection{Sources of noise}

\begin{table}
    \caption{Noise contributions from the instrument model.}
    \label{table:instrumentnoise}
    \centering
    \begin{tabular}{l l l }
       \hline\hline
        Symbol & Value & Description \\
        \hline
        $t_{\rm exp}$ & 550 s & Exposure time \\
        $n_{\rm dark}$ & 9.37 $\rm e^{-}/s$ & Dark current \\
        $\sigma_{\rm read}^2$ & 0.031 $\rm e^{-}$ & Read noise\\
        \noalign{\vskip 1pt}
        \hline
    \end{tabular}
\end{table}

We separately considered the sources of statistical noise that contribute to the measurement.
The detector noise is characterised by the dark current $n_{\rm dark}$ and read-noise variance $\sigma_{\rm read}^2$.  The total detector noise variance per pixel $\sigma^2_{\rm det}$ is then
\begin{equation}\label{eq:noise}
\sigma^2_{\rm det} = N_{\rm exp} \, \left( t_{\rm exp} \, n_{\rm dark} + \sigma_{\rm read}^2 \right)\, ,
\end{equation}
where $N_{\rm exp}$ is the number of exposures. The adopted pre-launch values, taken as averages over the focal plane, are listed in Table~\ref{table:instrumentnoise}. The source signal contributes to the noise through the Poisson uncertainty in the electron counts, $\sigma^2_{\rm signal}$.
Finally, the contribution from astrophysical diffuse foregrounds $\sigma^2_{\rm astro}$, including zodiacal light and scattered light from stars (or straylight), is added to the detector variance such that
\begin{equation}\label{eq:totnoise}
\sigma^2_{\rm total} = \sigma^2_{\rm det} + \sigma^2_{\rm signal} + \sigma^2_{\rm astro}\, .
\end{equation}
To model the zodiacal emission we used the \citet{Aldering2001} prescription. The straylight map was computed for \Euclid performance forecasts and made available in the Euclid Mission Database. As for the Milky Way extinction map, we used the \citet{PlanckDust2014} map.

\subsection{\fastspec}

The \fastspec simulation pipeline aims at rapidly simulating galaxy 1-dimensional spectra with the expected noise from \Euclid observations. It relies on the simplified instrument model described in the previous section to simulate the main steps involved in spectrum extraction from dispersed images. The input to the code is an incident noise-free spectrum and the bulge and disk morphological parameters needed to simulate the galaxy image.

The 2-dimensional spectrum image is constructed over a rectangular pixel grid using Eq.~\eqref{eq:convolution}. The object image and associated flux are modelled by a sum of 2-dimensional S\'ersic profiles, each one representing a morphological component of the galaxy. The morphological parameters, in particular the bulge and disk scale lengths, need to be provided for each object.

Once the dispersed image of the object is created, a random Gaussian noise realisation is drawn, independently for each pixel, based on the total variance given in Eq.~\eqref{eq:totnoise}. Depending on the required observational strategy, several noise realisations, and in turn 2-dimensional spectra, can be generated with varying dispersion directions. They correspond to different exposures of the same object. This allows us to mimic the \Euclid dithering pattern in spectroscopic observations. A 1-dimensional spectrum is then extracted from each individual exposure using Eq.~\eqref{eq:1dextraction}. The extraction window size, that is, the number of pixels over which the sum in Eq.~\eqref{eq:1dextraction} is performed, is a crucial parameter at this step that is fixed according to the size of the source (the optimised extraction based on the source profile is not implemented). Finally, multiple exposures are combined with the inverse variance-weighted mean to produce the final 1-dimensional spectrum. An associated error spectrum is also produced by propagating the pixel variance from the single exposures.

\fastspec is flexible enough to allow the simulation of different types of spectroscopic observations. In particular, the diffuse foreground noise level, instrument noise level, spectral resolution, number and length of exposures, and angles of the grism are parameters to be set. This pipeline is not intended to perform high-fidelity pixel-level simulations of \Euclid spectra. It assumes constant background noise for each spectrum and sensitivity across the focal plane, perfect decontamination of each individual object spectrum, and does not consider any residual error induced by spectrum trace location on the detector or by the 2-dimensional mapping of image pixels to wavelength. Moreover, \fastspec does not account for cosmic ray hits or detector persistence, which require masking in the real data. Rather, \fastspec gives an effective way of simulating observed 1-dimensional spectra, which quantitatively captures the Poisson contribution to the noise for individual sources.

 \begin{figure*}
 \includegraphics[width=\linewidth]{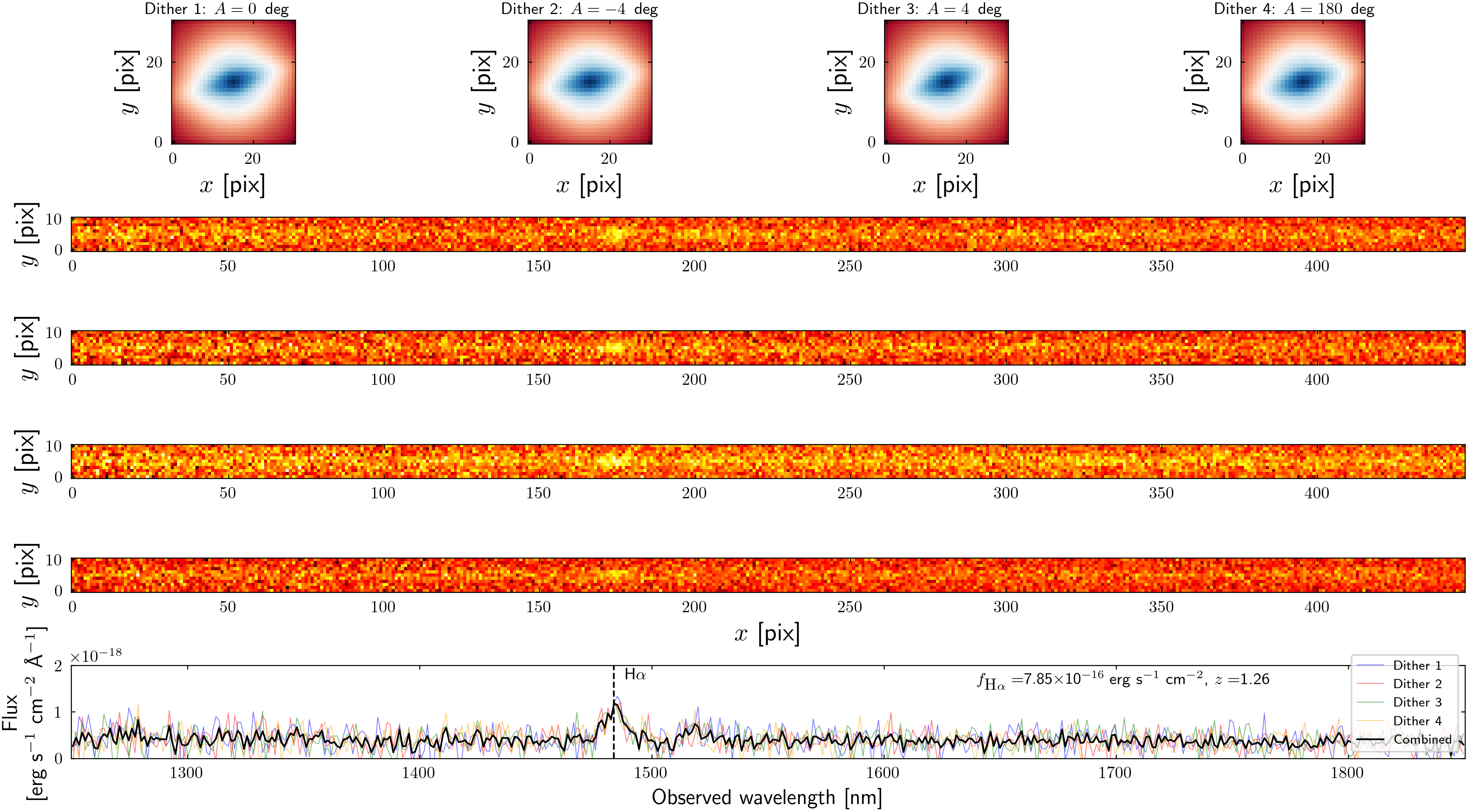}
  \caption{From top to bottom: simulated galaxy images in the different dithers (top panel), corresponding simulated 2-dimensional spectrograms (middle panels), extracted 1-dimensional spectral traces (bottom panel, coloured lines), and final combined 1-dimensional spectrum (bottom panel, black line) with \fastspec for an emission-line galaxy at $z=1.26$ with $f_{\Ha} = \num{7.85e-16}\,\ergcms$.
  \label{fig:fastspec}}
\end{figure*}

\subsection{\pypelid}

The \pypelid simulation pipeline is designed to efficiently simulate spectroscopic measurements over a large survey area. It comprises the spectral simulation, extraction, and redshift measurement steps, and the output consists of the signal-to-noise ratio (S/N) of emission lines, detection probability, and measured redshift value. \pypelid makes a number of simplifying assumptions with respect to \fastspec.

The \pypelid code runs in two stages. It first processes each spectroscopic exposure in the survey. For each exposure it determines which galaxies have emission lines that are in the wavelength range of the red grism. At least one emission line is required to land on a detector. The  emission lines that are simulated and have line fluxes specified in the input catalogue are \Ha, [\ion{N}{ii}]$\lambda\lambda~6550,6585$, [\ion{S}{ii}]$\lambda\lambda~6718, 6733$, \Hb, [\ion{O}{iii}]$\lambda\lambda~4960, 5008$, and [\ion{S}{iii}]$\lambda\lambda~9069, 9530$. The result of the processing is a table that contains for each galaxy and each emission line: the number of exposures; the sum of noise variance values; and the sum of the fluxes.

In the second stage, \pypelid computes a total S/N for detection based on the effective signal and noise values of each emission line. The determination of the total S/N is made based on the simulation of the 1-dimensional spectrum, so that it also accounts for the dependence on angular size and blending of emission lines.  The galaxy spectrum is simulated as a set of emission lines with line profiles constructed from galaxy morphological parameters and the optical PSF. The dispersed image is constructed as in Eq.~\eqref{eq:convolution}, considering only the flux in the discrete emission lines. In practice, the dispersed image is built by Monte Carlo photon shooting, as described in Appendix~\ref{app:pypelid}. The 1-dimensional spectrum is extracted and the model S/N is defined as
\begin{equation}\label{eq:SNR}
    \SN_{\rm model} = \sqrt{\sum_i \frac{f_i^2}{\sigma^2_i}} \, ,
\end{equation}
where the sum ranges over spectral pixels indexed by $i$. Only emission lines are simulated, thus the flux is non-zero only at
the location of the emission lines. The model S/N is used in the detection model described in Sect.~\ref{sec:bypass} to predict the probability that a given galaxy is measured correctly under its observational noise conditions.

\section{Euclid Wide Survey performance}

\subsection{Sample completeness, redshift purity, and interlopers}

We first introduce the key metrics for the spectroscopic survey performance in the context of \Euclid slitless spectroscopy, which include redshift purity and completeness with respect to an underlying `target' sample. The target sample represents a specific population of galaxies in the Universe that may be detectable by \Euclid. For our purposes, the  target sample is limited to galaxies in the redshift range 0.9 to 1.8, and with \Ha emission-line flux greater than $\num{2e-16}\,\ergcms$. We note that these criteria are based on design requirements. The EDS fields are expected to provide a census of the galaxy population that will allow us to tune the specifications of the target sample for galaxy clustering analysis.

The sample completeness $C$ is the fraction of the target sample that is contained in the selected catalogue and measured with the correct redshift,
\begin{equation}
    C := \frac{N_{\rm target\, \&\, selected \, \&\, correct}}{N_{\rm target}}\, .
\end{equation}
The redshift purity $P$ is the fraction of the selected sample that is measured with the correct redshift,
\begin{equation}
    P := \frac{N_{\rm selected\, \&\, correct}}{N_{\rm selected}}\, ,
\end{equation}
where we adopt the following criterion for correct redshift measurements motivated by the design requirements,
 \begin{equation}\label{eq:zerr}
     \frac{\left|z_{\rm meas}-z_{\rm true}\right|}{1+z_{\rm true}}  <  \num{3e-3} \, .
 \end{equation}
The normalisation by $1+z_{\rm true}$ is used to be consistent with \Euclid design documents. However, it is worth noting that the statistical error on redshift measurements does not necessarily scale in this way.

The impact of impurity on galaxy clustering statistics depends on the type and fraction of interloper redshifts \citep{EP-Risso}. We consider two classes of interlopers: line interlopers, which arise from the misidentification of a spectral feature in the measured spectrum with a different spectral feature; and noise interlopers, which arise from the misidentification of a spurious noise feature with a line. Line interlopers have a systematically biased redshift measurement due to the misidentification of the rest-frame wavelength as $\lambda_{\rm rest, wrong}$, when the true wavelength is $\lambda_{\rm rest, true}$, such that
\begin{equation}\label{eq:deltaz}
    \frac{1+z_{\rm true}}{1+z_{\rm meas}} = \frac{\lambda_{\rm rest, wrong}}{\lambda_{\rm rest, true}}\,,
\end{equation}
and
\begin{equation}
    \Delta z := \frac{z_{\rm meas} - z_{\rm true}}{1+z_{\rm true}} = \frac{\lambda_{\rm rest, true}}{\lambda_{\rm rest, wrong}} - 1\,.
\end{equation}
Noise interlopers instead do not have strong spectral line detections and the redshift measurement is the result of spurious noise features. In this case, we consider the measured redshift to be uncorrelated with the true redshift, since we have evidence of this in early data.

\subsection{\fastspec +~SPE simulation of the EWS}

To simulate the reference EWS we used the EL-COSMOS catalogue described in Sect.~\ref{sec:elcosmos}. The spectra were simulated with \fastspec in an EWS configuration considering galaxies with three or more exposures across the entire wavelength range. We limited the sample to galaxies with $\HE<24$. To improve statistics the catalogue was replicated five times. Each replicated galaxy was assigned a perturbed redshift with shift $\Delta z$ drawn from a Gaussian distribution with standard deviation 0.02. The sky coordinates were not used in the analysis. The background noise was drawn from a distribution between 1 and 3.74 $\mathrm{e^-\,pix^{-1}\,s^{-1}}$, as expected in the EWS.  Stars ($\texttt{type} = 1$ in the EL-COSMOS catalogue) were removed from our analysis. We expect that stars will be efficiently filtered out of the catalogue using photometric and morphological criteria, and this point will be studied on real data.  The final sample (counting replicated galaxies) used for the statistical studies presented in the following sections consists of approximately one million galaxies. Since each spectrum was simulated with an independent background noise, each realisation can be considered as statistically independent. Example EWS spectra simulated with \fastspec are shown in Fig.~\ref{fig:fastspec}. The simulated, combined 1-dimensional spectra were eventually analysed with the OU-SPE redshift measurement pipeline \citep{Q1-TP007}.

\subsection{Purity and completeness}

\begin{table}
    \caption{Redshift purity and sample completeness for selections defined on EL-COSMOS \fastspec simulations.}
    \label{table:puco}
    \centering
    \begin{tabular}{l l l c c}
       \hline\hline
       \noalign{\vskip 1pt}
        Selection & Criteria & $P$ & $C$ \\
        \hline
        \noalign{\vskip 1pt}
        Baseline & ${\tt spe\_z\_prob}>0.97$ &  0.74 & 0.45 \\
        Colour I & SVC \& ${\tt spe\_z\_prob}>0.96$ & 0.83 & 0.45 \\
        Colour II & SVC \& ${\tt spe\_z\_prob}>0.94$ & 0.78 & 0.49 \\
        \hline
    \end{tabular}
\end{table}

 \begin{figure}
    \includegraphics[width=3.5in]{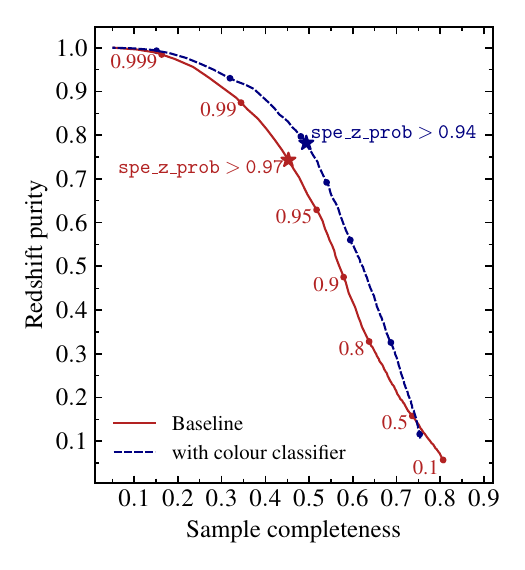}
    \caption{Redshift purity and sample completeness of the OU-SPE redshift measurements on EL-COSMOS \fastspec simulations. The curve is defined by varying the redshift measurement quality threshold $\tt spe\_z\_prob$ from 0 to 1. The points along the curves mark the labelled values. The star symbol marks the point that maximises the product of purity and completeness at the threshold $0.97$ for the baseline selection and $0.94$ with the colour selection using \Euclid onboard photometry. \label{fig:puco}}
\end{figure}

After producing the redshift measurements it is necessary to select a reliable sample useful for galaxy clustering analysis. For this, the OU-SPE pipeline outputs the redshift measurement probability \texttt{spe\_z\_prob}, which quantifies the quality of the redshift measurement. The quantity \texttt{spe\_z\_prob} is computed as the integral of the redshift probability distribution function in a window around the most probable redshift \citep{Q1-TP007}. 

To determine the best selection for galaxy clustering we used the redshift purity and sample completeness metrics. We computed the purity and completeness over a range of \texttt{spe\_z\_prob} thresholds. This produced a curve in the purity--completeness plane shown in Fig.~\ref{fig:puco}. We used the product of purity and completeness as the metric to optimise the selection. We found that the product is maximised with the threshold $\texttt{spe\_z\_prob} > 0.97$ giving a purity of $P=0.74$ and completeness of $C=0.45$ listed in Table~\ref{table:puco}. We adopted this configuration as our baseline sample. 

As studied in \citet{Cagliari24}, we also considered a photometric selection applied to the sample to identify the sources that are likely to be detectable \Ha emitters with $f_{\Ha}>\num{2e-16}\,\ergcms$ in the target redshift range. This selection is based only on the \Euclid \IE, \YE, \JE, and \HE photometry. We used a support vector classifier (SVC) with a radial basis function implemented in the Python \texttt{scikit-learn} library \citep{scikit-learn}. The training set for the classifier was constructed from the EL-COSMOS catalogue using a balanced set with two classes: targets that were correctly measured, and non-targets. The training set consisted of about 5\% of the total catalogue\footnote{In the case of real data, a machine learning classifier can be robustly trained on single-pass reductions of the Euclid Deep Fields and applied on the EWS.}. With the additional selection based on the \Euclid photometric bands, we found that at completeness $C=0.45$, the purity increased to $P=0.83$, referred to as the Colour I selection in Table~\ref{table:puco}. The point that optimises the product of purity and completeness is at $P=0.78$ and $C=0.49$, with the redshift measurement quality threshold of $\texttt{spe\_z\_prob} > 0.94$, referred to as the Colour II selection. It is important to note that the optimal threshold on \texttt{spe\_z\_prob} depends on the particular data set and selection under consideration, and differs in real data, as in the Q1 data release, for which a higher threshold was used \citep{Q1-TP007}.

\begin{figure*}
    \includegraphics[width=3.3in]{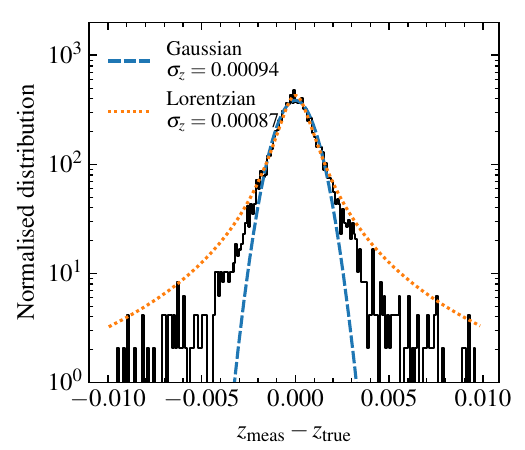}
    \includegraphics[width=3.8in]{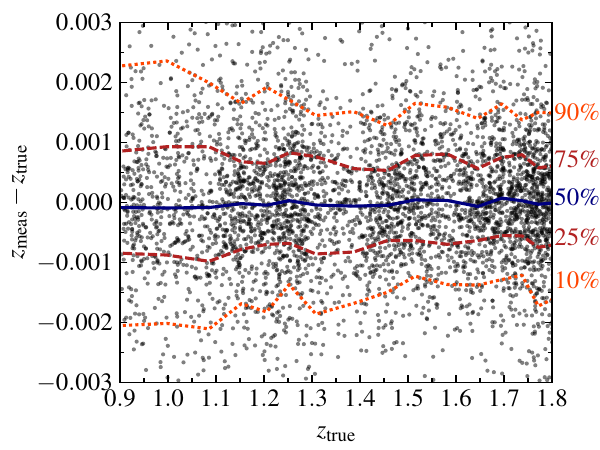}
    \caption{\emph{Left}: random redshift measurement error distribution assessed from the \fastspec simulations processed by OU-SPE. The dashed and dotted curves show the Gaussian ($\sigma_z=\num{9.4e-4}$) and Lorentzian model ($\sigma_z=\num{8.7e-4}$) best fits, respectively. \emph{Right}: trend of the redshift measurement error with redshift. The measurement error is approximately independent of redshift, and is not proportional to $1+z_{\rm true}$. From the bottom to the top, the lines mark respectively the 10th, 25th, 50th, 75th, and 90th percentiles in adaptive redshift bins that contain $300$ galaxies. \label{fig:randomzerror}
    }
\end{figure*}

\subsection{Redshift error}\label{sec:redshifterror}

 \begin{figure}
     \includegraphics[width=3.5in]{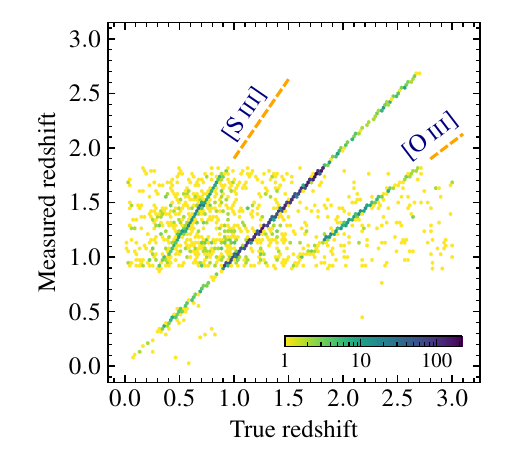}
     \includegraphics[width=3.5in]{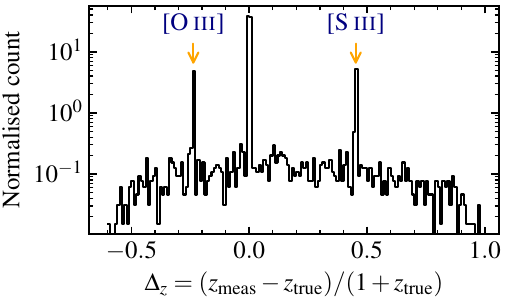}
    \caption{\emph{Top}: distribution of true and measured redshift, with the density of points indicated by the colour scale. Correct measurements fall on the diagonal while line interlopers fall along tracks with specific slopes. Noise interlopers are predominantly distributed over the prior redshift range 0.9 to 1.8. \emph{Bottom}:  interloper distribution, summarised by the histogram of redshift error $\Delta_z$, in which the line interlopers are projected into distinct peaks. \label{fig:interlopers}}
\end{figure}

\begin{table}
    \caption{Interloper fractions for EL-COSMOS \fastspec simulations with the baseline selection.}
    \label{table:fractions}
    \centering
    \begin{tabular}{l l l c c}
       \hline\hline
       \noalign{\vskip 1pt}
        Redshift  & [\ion{O}{iii}] & [\ion{S}{iii}] & Noise\\
        \hline
        \noalign{\vskip 1pt}
0.95 & 0.009 & 0.031 & 0.325 \\
1.05 & 0.016 & 0.029 & 0.354 \\
1.15 & 0.057 & 0.041 & 0.220 \\
1.25 & 0.074 & 0.057 & 0.159 \\
1.35 & 0.122 & 0.104 & 0.239 \\
1.45 & 0.075 & 0.138 & 0.154 \\
1.55 & 0.038 & 0.056 & 0.169 \\
1.65 & 0.017 & 0.040 & 0.127 \\
1.75 & 0.012 & 0.015 & 0.072 \\
        \hline
    \end{tabular}
\end{table}

\begin{figure}
    \includegraphics[width=3.5in]{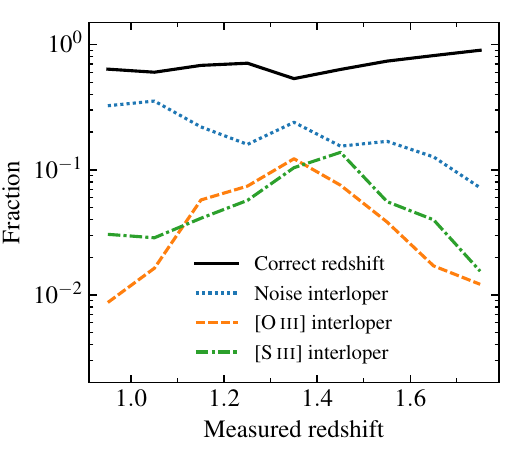}
    \caption{Redshift dependence of the interloper fractions, including line interlopers arising from [\ion{O}{iii}], [\ion{S}{iii}], and noise interlopers, listed in Table~\ref{table:fractions}.
    \label{fig:fractions}
    }
\end{figure}

We analysed the redshift measurement error for the baseline selection ($\texttt{spe\_z\_prob} > 0.97$) in the bypass EWS simulation. Figure~\ref{fig:randomzerror} shows the distribution of the redshift error $z_{\rm meas} - z_{\rm true}$ in the left panel, and a scatter plot showing its trend with $z_{\rm true}$ in the right panel. We found that the distribution has extended tails, but the core is well fitted by a Gaussian distribution with $\sigma_z = \num{9.4e-4}$. The fit with a Lorentzian distribution is overplotted and the amplitude in the tails falls between the best-fitting Gaussian and Lorentzian distributions. The right panel demonstrates that the redshift error does not have a significant dependence on redshift. The redshift error in slitless spectroscopy depends on multiple factors, including the number of spectral features used in the redshift estimate and the angular size of the galaxy. The redshift error trends in simulated EWS and EDS spectra were studied in detail in \citet{EP-Quai}.

Figure~\ref{fig:interlopers} shows the full distribution of redshift errors, including catastrophic ones that are classified as redshift interlopers. The majority of redshift measurements are in the range 0.9 to 1.8, due to the prior used in the template fitting that gives more weight to redshift solutions with the \Ha line \citep{Q1-TP007}. Good redshift measurements fall along the one-to-one diagonal. The redshift measurements that appear randomly distributed are the so-called noise interlopers in which a noise feature in the spectrum was misidentified as a spectral feature. The two off-diagonal tracks arise from line misidentifications in which a low-redshift galaxy has the [\ion{S}{iii}] emission line mistaken for \Ha, or a high-redshift galaxy has [\ion{O}{iii}] mistaken for \Ha. The 2-dimensional distribution of redshift errors can be condensed into the histogram of $\Delta_z$, shown in the bottom panel of Fig.~\ref{fig:interlopers}. Line interlopers map to specific $\Delta_z$ values, while noise interlopers populate a broad distribution. 

The redshift dependence of the interloper fractions is plotted in Fig.~\ref{fig:fractions} and listed in Table~\ref{table:fractions}. The trends with redshift are not intrinsic properties of the data. They are determined by the wavelengths of spectral features (e.g. \Ha and [\ion{O}{iii}] wavelengths), and the template set and priors used in the redshift fitting routine. We found that the noise interloper fraction tends to decrease with redshift, while the line interlopers peak at redshift 1.3--1.5.

\section{Bypass detection prescriptions}

 \begin{figure*}
  \includegraphics{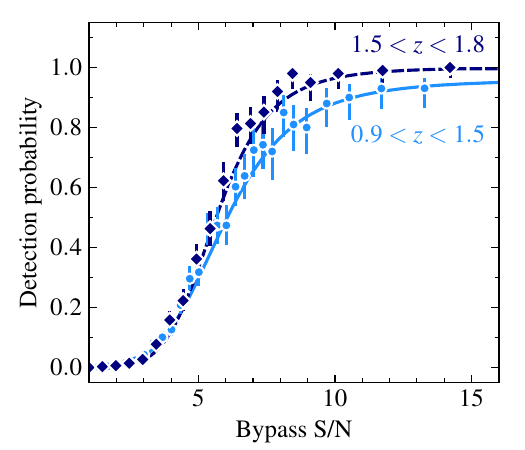}
  \includegraphics{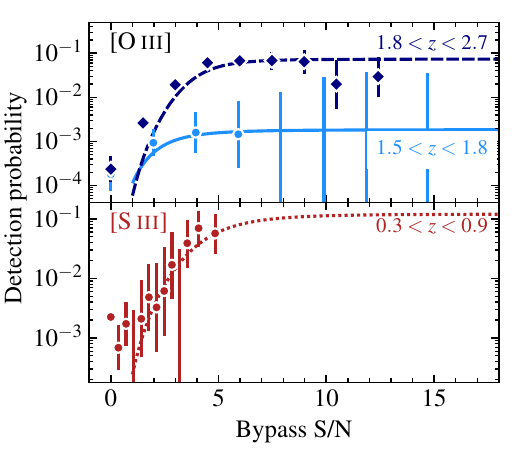}
  \caption{\emph{Left}: detection probability as a function of S/N from \fastspec simulations analysed with OU-SPE in two redshift bins, $0.9<z<1.5$ and $1.5<z<1.8$. The curves show the best-fitting model. \emph{Right}: detection probability of [\ion{O}{iii}] and [\ion{S}{iii}] emission lines mistaken for \Ha, which enter as interlopers. The indicated redshift bins refer to the true redshifts of the interloper samples.
  \label{fig:detprob}}
\end{figure*}

\subsection{S/N detection model} 
\label{sec:bypass}
We developed a detection model to describe the process of spectral extraction and redshift measurement of a target galaxy. The total S/N of the observed emission lines, defined in Eq.~\eqref{eq:SNR}, is the natural quantity to characterise the process.
A source is considered to be detected if it passes the sample selection criteria and has a redshift measurement error below the specified threshold of $|\Delta_z| < 0.003$ in Eq.~\eqref{eq:zerr}. This model can be extended to redshift interlopers by using the systematically shifted redshift given by Eq.~\eqref{eq:deltaz} as the reference. The detection model is parameterised with a sigmoid function of the form
\begin{equation}
    p(\SN) = \frac{c}{1 + (\SN/\SN_0)^{-b}}\, ,
\end{equation}
where the three parameters $c$, $\SN_0$, and $b$ have to be calibrated. We fitted them on the redshift measurements from the EL-COSMOS \fastspec +~SPE simulation in different redshift bins. The first step was to apply the selection criteria to the simulated measurements, i.e. ${\tt spe\_z\_prob}>0.97$. We then split the selected sample into four classes: correct redshift measurements, [\ion{O}{iii}] line interlopers, [\ion{S}{iii}] line interlopers, and noise interlopers.  We fitted a model to each class and the best-fitting model gives the probability that a source with a given S/N and redshift is detected in the given class.

The fit to correct redshift measurements is shown in the left panel of Fig.~\ref{fig:detprob}. The sample was divided into two redshift bins, $0.9<z<1.5$ and $1.5<z<1.8$, which were fitted separately. The split at $z=1.5$ corresponds to the redshift at which the [\ion{O}{iii}] doublet enters the wavelength range of the red grism. We found that the detection probability measured in the simulated data is higher at $z>1.5$, which can be understood since both line complexes \Ha and [\ion{O}{iii}] are detectable in that range, while only \Ha is detectable at $z<1.5$.
The fits to the line interloper classes, [\ion{O}{iii}] and [\ion{S}{iii}], are shown in the right panel of Fig.~\ref{fig:detprob}. The [\ion{O}{iii}] class was split into two redshift bins, $1.5<z<1.8$ and $1.8<z<2.7$. The probability of detecting [\ion{O}{iii}] as an interloper is approximately two orders of magnitude lower in the first redshift bin, since in that range both \Ha and [\ion{O}{iii}] are detectable, and thus the likelihood of having a line interloper is significantly lower. For [\ion{S}{iii}] interlopers, we considered a single redshift range $0.3 < z < 0.9$ for the model. It is worth pointing out that the sigmoid model asymptotes to a constant $c$ at high S/N, which makes sense for correct redshift measurements, but may not reflect the genuine distribution of line interlopers. In the current simulations we do not have sufficient statistics to probe the interloper distribution in the high S/N regime, so this aspect is left to further study.

Galaxies that passed the selection but were not labelled as correct redshift measurements or line interlopers were classified as noise interlopers. By definition noise interlopers do not have strong emission lines, and thus we cannot use S/N to parameterise the distribution. Instead we chose to assign a constant probability that a source enters as a noise interloper. In the simulated sample, this resulted in a probability of $p_{\rm noise} = 0.023$, that is, approximately 2\% of the simulated galaxies entered the sample as noise interlopers. Since the target sample is a small fraction of the observed galaxies, the 2\% leads to a contribution of up to 35\% in the noise interloper fraction, as tabulated in Table~\ref{table:fractions} and shown in Fig.~\ref{fig:fractions}. A prescription is also needed to assign the measured redshift to a noise interloper. We achieved this by randomly sampling from the measured redshift distribution.

The detection model framework successfully reproduced both the input interloper fractions and their redshift distributions by specifying only the S/N and redshift, as illustrated in Appendix~\ref{app:trends}. We further examined trends related to secondary intrinsic galaxy properties, finding that these are also well reproduced, as detailed in the same appendix.

\subsection{Tabulated detection probability}

The S/N computation implemented in \pypelid\ is based on simulated spectra accounting for both line fluxes and galaxy profiles. However, profile information is not always available, in particular, it was absent from the HOD-based galaxy mock catalogues of \citet{EP-Monaco1}, which provided only \Ha line flux values. This led us to marginalise the detection model over the galaxy profile parameters, reducing its dependence to \Ha flux alone. In this case, we constructed tabulated models mapping \Ha flux, redshift, and Galactic extinction to S/N, which were subsequently applied in \citet[][section~4.3]{EP-Monaco2}. 

Differences in the distributions of galaxy properties between EL-COSMOS and Flagship can lead to different detection probabilities. We minimised this effect by matching the distribution of \Ha flux in the EL-COSMOS catalogue to the distribution in Flagship,  as described in Sect. \ref{sec:elcosmos} and Fig. \ref{fig:nz}. Secondary dependencies, including morphology and surface brightness, also affect the detection probability through S/N, which highlights the importance of the accuracy of the mock galaxy catalogues.

\subsection{Redshift error model}

For specific studies, it is convenient to decouple angular selection effects, which depend on the S/N, from line-of-sight redshift errors. We therefore developed a complementary approach to model redshift errors, including interloper contamination, which can be directly used to assign realistic redshift errors to mock galaxy catalogues. We followed a supervised strategy in which a parametric model was fitted to the conditional probability $\mathcal{P}(z_{\rm meas} | z_{\rm true}, f_{\Ha})$ estimated from the EL-COSMOS sample. While the S/N detection model presented in Sect. \ref{sec:bypass} adopts a specific selection, the conditional probability distribution presented here characterises the full galaxy distribution as a function of \Ha flux, and is thus more general.

We first constructed a fine kernel density estimate of the conditional probability from the EL-COSMOS sample as
\begin{equation}
    \hat{\mathcal{P}}(z_{\rm meas} | z_{\rm true}, f_{\Ha}) = \frac{\hat{\mathcal{P}}(z_{\rm meas}, z_{\rm true}, f_{\Ha})}{\hat{\mathcal{P}}(z_{\rm true}, f_{\Ha})}\, ,
\end{equation}
where $\hat{\mathcal{P}}(z_{\rm true}, f_{\Ha})$ is the marginal distribution of 
$(z_{\rm true}, f_{\Ha})$ and $\hat{\mathcal{P}}(z_{\rm meas}, z_{\rm true}, f_{\Ha})$ is the corresponding joint distribution. We then fitted this estimate in bins of $z_{\rm true}$ and $f_{\Ha}$ with the following parametric model
\begin{align}
    \mathcal{P}&(z_{\rm meas} | z_{\rm true}, f_{\Ha}) = f_{\rm c} (z_{\rm true}, f_{\Ha}) \, \mathcal{N}(z_{\rm meas} - z_{\rm true}; \sigma_z) \nonumber \\
    &+ \sum_{i=1}^{N_{\rm li}} \, f_{{\rm \ell}, i} (z_{\rm true}, f_{\Ha}) \, \mathcal{N}(z_{\rm meas} - z_{{\rm interloper},i}; \sigma_z) \nonumber \\
    &+ \left[1- f_{\rm c}(z_{\rm true}, f_{\Ha}) - \sum_{i=1}^{N_{\rm li}} f_{{\rm \ell}, i}(z_{\rm true}, f_{\Ha})\right]  \, \mathcal{B}(z_{\rm meas}; \theta) \,,
\end{align}
where $f_{\rm c}(z_{\rm true}, f_{\Ha})$ and $f_{{\rm \ell},i}(z_{\rm true}, f_{\Ha})$ are the fractions of correctly identified and line-interloper galaxies with emission line $i$, respectively. Here $\mathcal{N}(x;\sigma_z)$ denotes the normal distribution with standard deviation $\sigma_z$, $N_{\rm li}$ is the number of distinct line-interloper types. We define the systematically wrong redshift of the interloper type $z_{{\rm interloper},i}=(1+z_{\rm true})R_i - 1$, where $R_i = \lambda_{\ha} / \lambda_i$ is the ratio of the rest-frame \Ha wavelength to that of the misidentified emission line $\lambda_i$. The last term, $\mathcal{B}(z_{\rm meas}; \theta)$, is a positive polynomial distribution with parameters $\theta$, which captures the residual contribution from noise interlopers. The three components of the model thus account for correct redshift, line interlopers at specific wavelength ratios, and the noise interloper smooth background, respectively. In practice, the redshift error was held fixed for both correct and line-interloper populations in each fitting bin, and only the fractions $f_{\rm c}$ and $f_{{\rm \ell},i}$ were varied. We also neglected the dependence of the redshift error on \Ha flux and S/N, and used the fixed value for the standard deviation $\sigma_z = 10^{-3}$ (similar to the value measured in the selected sample in Sect. \ref{sec:redshifterror}). The best-fitting interloper fractions were subsequently interpolated continuously over the $(z_{\rm true}, f_{\Ha})$ plane to yield the final model for $\mathcal{P}(z_{\rm meas} | z_{\rm true}, f_{\Ha})$.

\begin{figure*}
    \includegraphics[width=\linewidth]{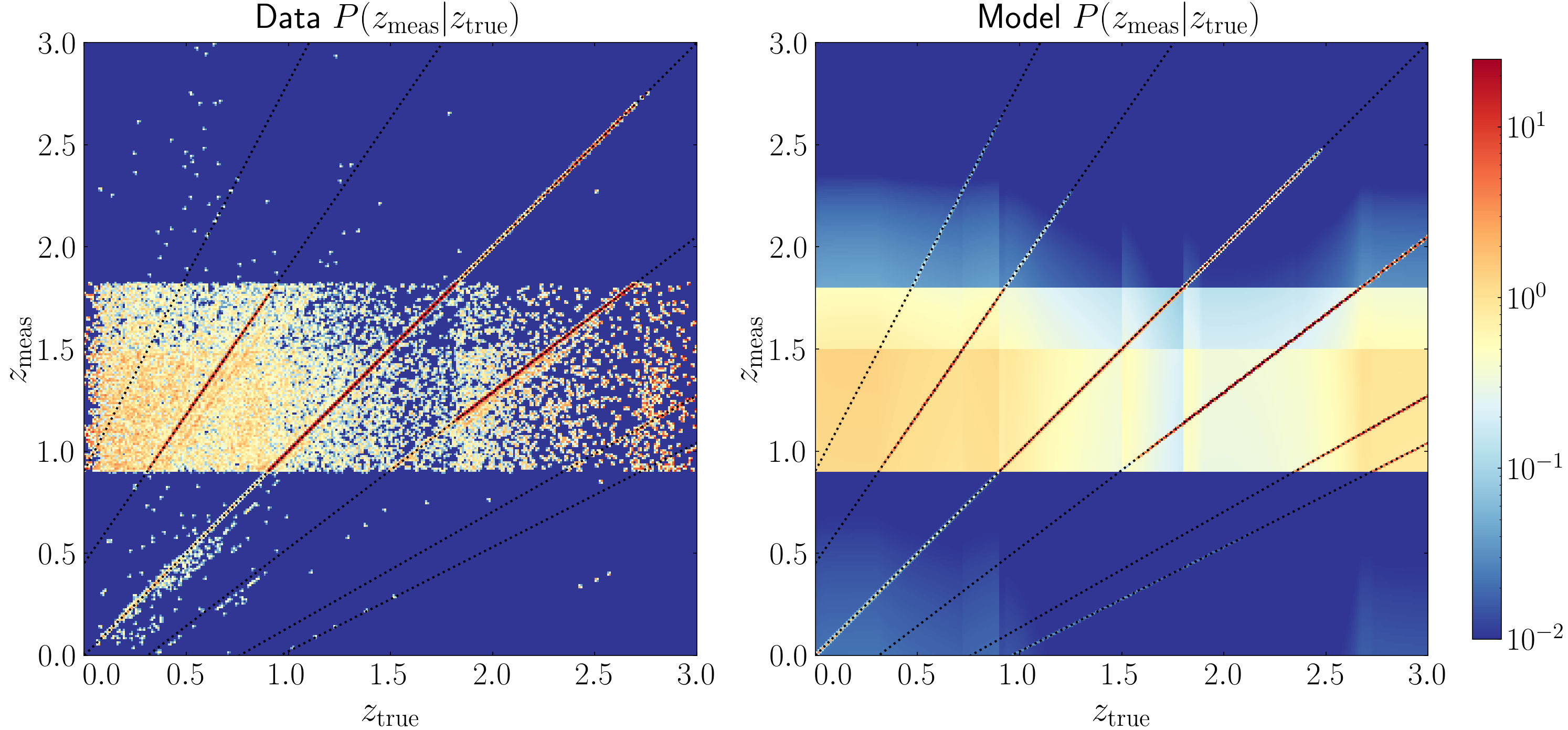}
     \caption{Conditional probability $\mathcal{P}(z_{\rm meas} | z_{\rm true})$ for galaxies 
with \Ha flux $f_{\Ha} \ge 2\times10^{-16}\,\ergcms$, as estimated 
from the EL-COSMOS catalogue (left) and predicted by our parametric model (right). The 
bright diagonal ridge traces the population of correctly identified \Ha emitters, 
while the secondary ridges correspond to line-interloper populations at specific wavelength ratios $R_i = \lambda_{\Ha}/\lambda_i$, overlaid as dotted lines. Both panels share the same logarithmic colour scale.\label{fig:condProb}}
\end{figure*}

The two panels of Fig.~\ref{fig:condProb} illustrate the quality of the model fit for galaxies with $f_{\Ha} \ge 2\times10^{-16}\,\ergcms$, which is the nominal flux limit of the EWS. The left panel shows the conditional probability $\hat{\mathcal{P}}(z_{\rm meas} | z_{\rm true})$ estimated directly from the EL-COSMOS catalogue, while the right panel shows the prediction of our parametric model. The dominant feature in both panels is the bright diagonal ridge along $z_{\rm meas} = z_{\rm true}$, corresponding to the population of correctly identified \Ha emitters with fraction $f_{\rm c}$. The narrowness of this ridge reflects the small redshift uncertainty of $\sigma_z = 10^{-3}$ assumed for correct identifications. Away from the diagonal, a series of secondary ridges are visible, each corresponding to a distinct line-interloper population. These ridges follow the loci $z_{\rm meas} = z_{\rm true} - (1+z_{\rm true})(1-R_i)$, which are overlaid as dotted lines for the four most significant $R_i$ values. The relative brightness of these ridges with respect to the main diagonal encodes the interloper fractions $f_{{\rm \ell},i}$, which vary smoothly with $z_{\rm true}$. A residual diffuse background, most visible at low redshift, corresponds to the noise interloper component modelled by a polynomial $\mathcal{B}(z_{\rm meas}; \theta)$. The right panel reproduces all these features with good fidelity. The positions, widths, and relative amplitudes of the main diagonal and the interloper ridges are well captured by the model across the redshift range of interest $0.9 < z_{\rm meas} < 1.8$. Minor discrepancies at low redshift are expected, given the flexible but low-order polynomial form adopted for noise interlopers. Overall, the comparison confirms that the parametric model provides an accurate and compact description of the redshift error distribution, suitable for application to large mock galaxy catalogues. This model was applied to assign realistic redshift errors to the mock catalogues used in \citet{EP-Risso} and \citelee to study the impact of redshift interlopers on the galaxy 2-point correlation function and power spectrum.

\section{Simulation production for galaxy clustering quality assessment}

 \begin{figure*}
  \includegraphics[width=7in]{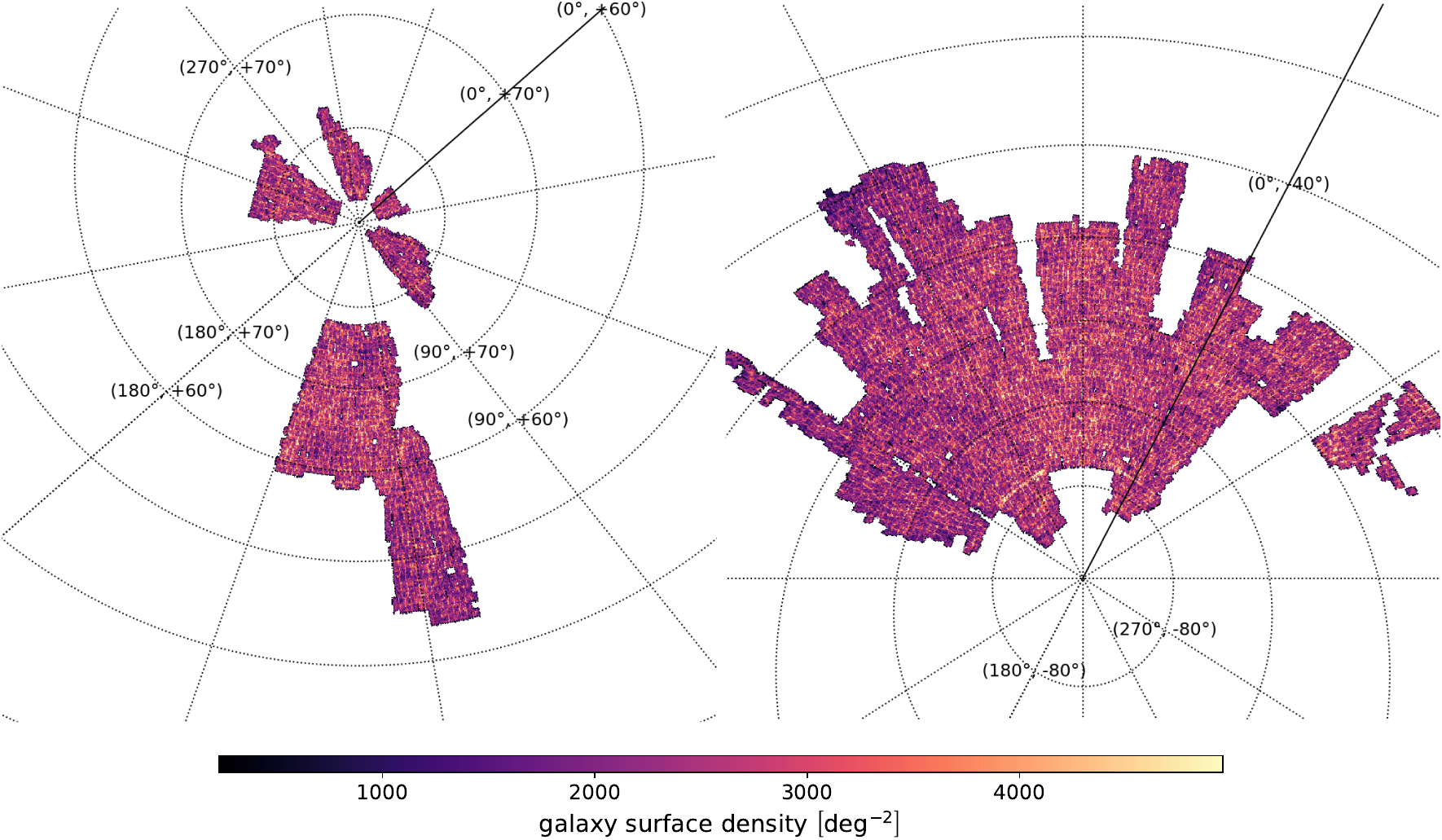}
  \caption{Map of projected galaxy density in the DR1 survey simulated with the Flagship mock galaxy catalogue and \pypelid. Systematic variations in number density are caused by modulations in signal-to-noise ratio induced by the dither pattern of exposures, the zodiacal light background, scattered starlight, and Milky Way extinction.
  \label{fig:pypelidmap}}
\end{figure*}

 \begin{figure*}
  \includegraphics[width=7in]{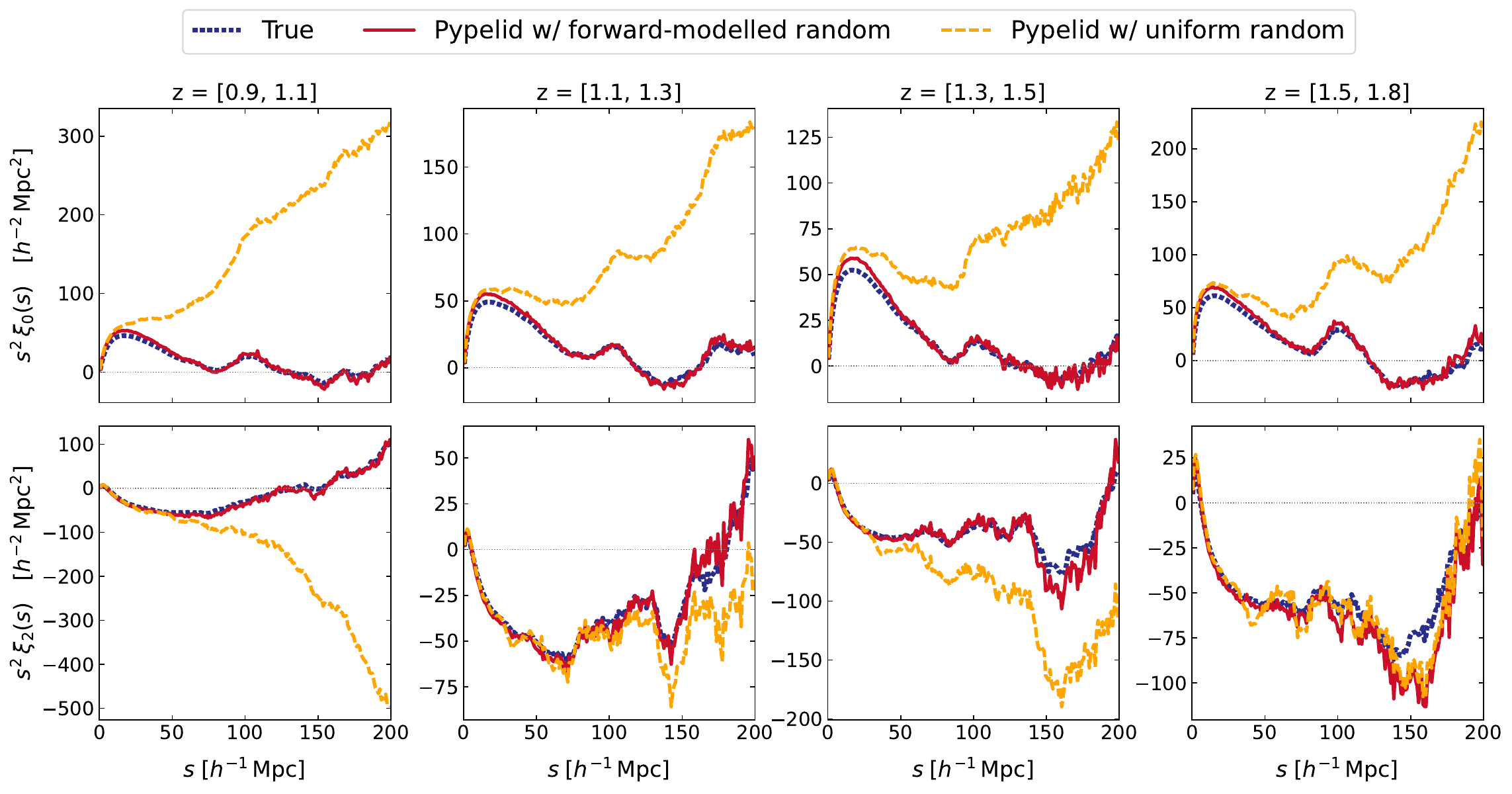}
  \caption{Two-point correlation function monopoles (\emph{top panels}) and quadrupoles (\emph{bottom panels}) in the simulated DR1 survey based on the Flagship mock galaxy catalogue, in the four redshift intervals, $[0.9, 1.1]$, $[1.1, 1.3]$, $[1.3, 1.5]$, and $[1.5, 1.8]$. The reference truth (blue dotted curve) is measured from the \Ha flux-limited sample ($f_{\Ha}>10^{-16}\,\ergcms$) without observational systematics. Analysing the \texttt{Pypelid} simulation with a uniform random catalogue yields excess  power in the monopole and quadrupole correlation functions (orange dashed curve). These systematic errors can be corrected by using forward modelling to imprint the selection effects in the random catalogue (red solid curve). The corrected monopole measurement has a different amplitude by 10--20\% with respect to the reference, due to the change in galaxy bias of the sample after effectively changing the flux limit due to the S/N detection process modelled with \pypelid.
  \label{fig:pypelidcf}}
\end{figure*}

As a proof of concept, we ran a \pypelid simulation of the Euclid Data Release 1 (DR1) survey. The input catalogue was constructed from the Flagship mock galaxy catalogue (Sect.~\ref{sec:flagship}). Figure~\ref{fig:pypelidmap} shows the projected galaxy surface density of the simulated DR1 footprint in ecliptic coordinates, for the north ($\beta \geq 43\degr$) and south ($\beta \leq -38\degr$) fields separately. The imprint of the survey footprint and exposure dither pattern is clearly visible, with regions of higher exposure count exhibiting a higher observed galaxy surface density. The map reveals large-scale gradients in density driven by spatially varying noise levels: the zodiacal background and scattered starlight modulate the S/N across the footprint, as does Milky Way extinction, which attenuates both the 
\Ha line flux and the continuum of background galaxies. The southern field, which covers a significantly larger and more contiguous area, shows a rich structure of density variations at the field boundaries and along tile edges, 
reflecting the complex tiling pattern of the EWS. These features illustrate that a realistic simulation of the \Euclid selection function must account for the full spatial complexity of the observing strategy, rather than 
assuming a uniform depth across the footprint.

We computed the 2-point correlation function monopole and quadrupole of 
the simulation after applying the S/N-based detection model. The monopole and quadrupole moments in four redshift intervals, $[0.9, 1.1]$, $[1.1, 1.3]$, $[1.3, 1.5]$, and $[1.5, 1.8]$, are presented in Fig.~\ref{fig:pypelidcf}. The ground truth reference was constructed from the flux-limited mock galaxy catalogue without any observational S/N selection. We considered two measurement configurations that differ in the treatment of the random catalogue used to estimate the correlation function. In the first configuration, a uniform random 
catalogue following only the survey footprint was used, without accounting for systematic angular and radial variations in density induced by the selection function \citep[referred to as the `ad hoc random' in][]{EP-Monaco2}. In the second configuration, a random catalogue was used that reproduces these systematic variations through a forward-modelling procedure: mock galaxies were drawn from the Flagship mock catalogue, placed at random positions within the survey footprint, and \pypelid was run on this random catalogue to imprint the same variation in S/N present in the data simulation. While we only considered the 2-point correlation function here, the impact of angular selection effects on the power spectrum was presented in \cite{EP-Monaco2}.

The results shown in Fig.~\ref{fig:pypelidcf} demonstrate that the choice of random catalogue has a dramatic impact on the measured clustering statistics. When a uniform random catalogue is used (orange dashed curve), the monopole correlation function is strongly boosted at all scales and in all redshift bins, with the signal rising steeply towards large separations rather than following the expected BAO feature and turnover. The quadrupole is similarly contaminated, exhibiting a large spurious power that completely masks the true RSD signal, particularly in the $0.9 < z < 1.1$ bin where the effect is the most severe. These distortions arise because the uniform random catalogue fails to capture the angular and radial density gradients introduced by the spatially varying selection function, leading to a systematic mismatch between the data and random pair counts, which in turn propagates directly into the measured 2-point correlation function. In contrast, when the forward-modelled random catalogue is used (red solid curve), the recovered monopole and quadrupole are in excellent agreement with the ground truth in all four redshift bins, accurately reproducing both the shape and amplitude of the BAO feature in the monopole and the RSD anisotropy signature in the quadrupole. This demonstrates that the forward-modelled random catalogue effectively accounts for the systematic density variations induced by the survey strategy. 

It is worth noting that the correlation function recovered with the forward-modelled random catalogue has a different amplitude by 10--20\% with respect to the reference on scales below $60\,\hMpc$.  This is expected, since the S/N-based detection model   effectively changes the flux limit. The reference mock catalogue was limited at $f_{\Ha} > 10^{-16} \,\ergcms$. After applying the detection model, the effective flux limit depends on observational quantities such as galaxy profile, redshift, and sky background, which introduce  secondary modulations of the selected galaxy population. As a result, the \pypelid catalogue probes a different galaxy population than the flux-limited reference, with a correspondingly different effective galaxy bias, which manifests as a change in the small-scale clustering amplitude. The change of amplitude of the monopole and quadrupole is consistent with a change of bias, although a quantitative fit has not been carried out.  The systematic change in galaxy properties due to the flux limit can be mitigated, but not fully removed, by matching the mean flux distribution of the reference catalogue to the \pypelid catalogue. In real data, the spectroscopic detection and redshift measurement process naturally imposes a spatially varying limit on S/N, which must be accounted for when comparing to simulations.

This study highlights the critical importance of accounting for spatially varying selection effects when computing 2-point statistics in wide-field spectroscopic surveys such as \Euclid. A uniform random catalogue, even when properly masked to the survey footprint, is insufficient to mitigate the systematic biases introduced by depth variations across the sky. A random catalogue constructed with forward modelling can account for these observational selection effects.

\section{Discussion}

The bypass simulation framework developed for the \Euclid spectroscopic redshift survey encompasses two complementary tools, \fastspec and \pypelid, as well as calibrated prescriptions for the survey selection function, redshift errors, and redshift interloper populations. These tools were designed to reproduce the essential characteristics of the \Euclid slitless spectroscopic measurements at a computational cost that is orders of magnitude lower than full pixel-level simulations, enabling their application to the large mock catalogue ensembles required for galaxy clustering analyses.

\fastspec simulations applied to EL-COSMOS enable us to assess the redshift quality statistics, purity, completeness, and interloper fractions, which characterise the EWS. We demonstrated that the baseline selection applied to simulated EL-COSMOS spectra yields a redshift purity of 74\% and a sample completeness of 45\%. The inclusion of a photometric colour classifier based on \Euclid broadband photometry modestly improves these figures to 78\% and 49\%, respectively. These results are broadly consistent with expectations from pre-launch performance models and provide a quantitative foundation for systematic error budget assessments. The redshift error distribution obtained from \fastspec simulations is well described by a Gaussian core with width $\sigma_z = 9.4 \times 10^{-4}$, with extended non-Gaussian wings that fall between Gaussian and Lorentzian profiles. Importantly, we found that the redshift error does not show a significant trend with redshift over the range $0.9<z<1.8$, and is not proportional to $1+z$, a point of relevance for the modelling assumptions made in galaxy clustering analyses. The interloper population is dominated at low S/N by noise interlopers, while line interlopers from [\ion{O}{iii}] and [\ion{S}{iii}] misidentifications contribute at the level of a few to ten per cent, depending on redshift, with a characteristic redshift dependence that can be understood from the wavelength coverage of the red grism. The conditional redshift error model $\mathcal{P}(z_\mathrm{meas}|z_\mathrm{true})$ calibrated on those simulations provides a complementary prescription that decouples angular from line-of-sight selection effects, and was applied to assign realistic redshift errors to mock catalogues.

\pypelid extends this framework to the full survey footprint. It operates on catalogues, accounting for the exact dither sequence, varying noise levels, and the angular selection function of the survey. The output includes detection probabilities and redshift measurements for input mock galaxy catalogues. The detection model, parameterised as a sigmoid function of the bypass S/N, was calibrated against \fastspec+~SPE simulations and reproduces the detection rate for correct redshifts and interloper classes across the relevant redshift ranges. We further constructed a tabulated model, marginalised over galaxy properties, that maps \Ha flux to detection probability. This allowed the framework to be applied to mock catalogues that contain only \Ha flux information, as in halo occupation distribution mocks. 

We performed a \pypelid simulation of the DR1 survey using the Flagship mock galaxy catalogue. The resulting galaxy density map reveals the complex spatial structure of the selection function, with number density modulations driven by the dither pattern of exposures, the zodiacal background, scattered starlight, and Milky Way extinction. These features demonstrate that a realistic simulation of the \Euclid selection function must account for the full spatial complexity of the observing strategy. The impact of the selection function on 2-point clustering statistics was assessed by computing the monopole and quadrupole moments of the correlation function in four redshift bins over $0.9 < z < 1.8$. We showed that a forward-modelled random catalogue produced by running \pypelid on a set of randomly placed mock galaxies recovers the monopole and quadrupole in excellent agreement with the ground truth across all redshift bins, accurately reproducing both the BAO feature and the RSD anisotropic signature. Looking forward, the approach can be applied to real data with a calibrated bypass detection model that accurately reproduces the survey selection function to robustly recover galaxy clustering statistics.

Several simplifying assumptions are embedded in the current framework that should be borne in mind when interpreting its outputs. \fastspec assumes a spatially uniform background noise and sensitivity across the focal plane, and does not model residual contamination from overlapping spectra. Contamination in the extracted spectra from spectral overlaps and artefacts including detector persistence and 0$^{\rm th}$-order images led to spurious detections in early \Euclid data \citep{Q1-TP007}. A dedicated study of the impact of contamination from overlapping continuum sources was presented in \citet{EP-Passalacqua}.

As we approach Euclid DR1, an important next step is to validate the bypass model predictions against measurements on real data, updating the detection model calibration where necessary. Source injection simulations that add artificial galaxies to real data will provide essential input for calibration \citeppassa. The use of realistic simulations that account for artefacts in the data is particularly important for characterising the detection model of the EDS where the substantially increased exposure depth leads to a different S/N distribution and interloper composition.

Systematic uncertainties and artefacts present in the \Euclid data can be accounted for by adding dependencies to the bypass detection model. For instance, the detection model could account for the number of overlaps with bright sources to assign a probability of a spurious detection from residual contamination.   Fundamentally, the detection model calibration will depend on the features present in the training dataset, and whether they are simulated or derived from real observations. Due to the complexity of the measurement process, it will be natural to turn to machine-learning-based methods for generalisation of the detection model.

The bypass framework built on simplified but physically motivated instrument models provides an efficient and reliable route to characterising the \Euclid spectroscopic selection function at the level of precision required for galaxy clustering science. The bypass spectroscopic simulation framework, together with the associated detection model and redshift error prescriptions, constitutes an integral component of the \Euclid simulation infrastructure for systematic assessments \citep{EP-Monaco2}, and forms the basis for ongoing preparation of the cosmological analyses that are the primary science goal of the \Euclid mission.

The \pypelid and \fastspec codes are made available in the Euclid Consortium open-source software repository\footnote{\url{https://github.com/euclid-consortium}}.

\begin{acknowledgements}
The authors dedicate this work to the memory of Bianca Garilli, whose foundational contributions to Euclid spectroscopy and unwavering mentorship continue to guide our community. She led the way in developing the first end-to-end simulations of \Euclid spectroscopy. BRG, IR, and MM acknowledge support from MIUR, PRIN 2022 (grant 2022NY2ZRS~001). KM acknowledges the sponsorship of grant 12~-~EUCLID11~-~0004, under a contract with the National Aeronautics and Space Administration (80NM0018D0004) at the Jet Propulsion Laboratory, California Institute of Technology. We acknowledge support by the Italian Research Centre on High Performance Computing Big Data and Quantum Computing (ICSC), by the Fondazione ICSC, and by the PRIN 2022 PNRR project (code no. P202259YAF).
\AckEC
\AckCosmoHub\xspace
We acknowledge support from the ELSA project with grant agreement number 101135203 funded by the European Union. Views and opinions expressed are, however, those of the authors only and do not necessarily reflect those of the European Union or Innovate UK. Neither the European Union nor the granting authority can be held responsible for them. UK participation is funded through the UK Horizon guarantee scheme under Innovate UK grant 10093177.
\end{acknowledgements}

\bibliographystyle{aa}
\bibliography{Euclid,references}

\appendix

\section{Point-spread function model}\label{app:psf}

The instrumental point-spread function model assumes radial symmetry
with a two-component Gaussian mixture parameterised by scales
$s_1$, $s_2$ and the relative amplitude $A_{\rm PSF}$. The cumulative radial
profile is given by
\begin{eqnarray}\label{eq:psf}
   p_{\rm PSF}(<r) &=& A_{\rm PSF} \left[1 - \exp\left(-\frac{r^2}{2 s_1^2}\right)\right] \nonumber \\
   &&+ \left(1-A_{\rm PSF}\right)\left[1-\exp\left(-\frac{r^2}{2s_2^2}\right)\right]\,.
\end{eqnarray}

\section{Emission-line spectrum simulation in \pypelid}\label{app:pypelid}

This appendix describes the implementation of emission-line spectrum simulation in \pypelid. The latter only simulates the emission lines associated with input objects with no account for the continuum or absorption lines. The counts recorded on the detector are assumed to be the result of a Poisson process with an expectation value given for a monochromatic line flux $f_{\rm line}$ by
\begin{equation}\label{eq:linesig}
s_{\rm line} = \int_{\lambda_{\rm start}}^{\lambda_{\rm end}}R(\lambda)\, f_{\rm line} \, \deltaD(\lambda - \lambda_{\rm line}) \, \diff \lambda \, ,
\end{equation}
where $\deltaD$ is the Dirac delta function characterising the monochromatic emission line.
The spectrum image is constructed with the following steps.
\begin{enumerate}
  \item For an emission line with a given flux, generate a Poisson realisation $N$ of the number counts recorded on the detector with expectation given by Eq.~\eqref{eq:signal}.
  \vspace{2pt}
  \item Draw $N$ samples $(x_{\rm s}, y_{\rm s})$ from the 2-dimensional source profile using the procedure described in Appendix~\ref{app:profile}.
  \vspace{2pt}
  \item Draw $N$ samples $(\Delta x, \Delta y)$ from the 2-dimensional PSF profile.
  \vspace{2pt}
  \item The $x$ coordinate is defined along the dispersion direction and $y$ in the perpendicular one.  The pixel position of the line is $x_{\rm line}=(\lambda_{\rm line}-\lambda_0)/d_{\lambda} + x_0$. The samples $(x_{\rm s}, y_{\rm s})$ are convolved with the PSF and velocity dispersion to give final coordinate positions:
\begin{eqnarray}
  x_{\rm image} &=& x_{\rm line} + x_{\rm s} + \Delta x \, ,  \\
  y_{\rm image} &=& y_{\rm s} + \Delta y \, .
\end{eqnarray}
\item The extraction of the 1-dimensional spectrum is performed in the same way as \fastspec, with an option to sum over the extraction window, or consider a weight function.
\end{enumerate}

\section{Galaxy image profile}\label{app:profile}

The galaxy image profile projected on the sky is composed of bulge and disk
components. The intensity profile is given by
\begin{equation}
I_{\rm total}(r) = (1-f_{\rm bulge}) \, I_{\rm disk}(r) + f_{\rm bulge} \, I_{\rm bulge}(r)\, .
\end{equation}
The disk is represented by an exponential profile and the bulge by a de
Vaucouleurs profile. A coordinate transformation is then performed to model 
the ellipticity and orientation of the source. We define the minor-to-major 
axis ratio as $q=b/a$ and the position angle (north to east) $\phi$. After 
defining the Cartesian coordinates $x'=r \cos\theta$ and $y'=r \sin\theta$, 
the final transformed coordinates on the sky are
\begin{eqnarray}
x &=& x' \, \sqrt{q} \, \cos \phi  + y' \, \sqrt{q^{-1}} \, \sin \phi\, , \label{eq:transform1} \\
y &=& -x' \, \sqrt{q} \, \sin \phi + y' \, \sqrt{q^{-1}} \, \cos \phi \label{eq:transform2} \,.
\end{eqnarray}

The image of the galaxy is rendered by Monte Carlo sampling its profile. To draw $N$ samples of
coordinates $(x,y)$ from the combined bulge and disk galaxy profile we perform
the following steps.
\begin{enumerate}
  \item Define $N_{\rm bulge} = f_{\rm bulge} \, N$ and $N_{\rm disk} = N - N_{\rm bulge}$, and draw samples of $r$
  from the two profiles independently.
  \vspace{2pt}
  \item Draw $N$ samples of $\theta$ uniformly in $[0,2\pi)$.
  \vspace{2pt}
  \item Transform $(r,\theta)$ to $(x,y)$ following Eqs.~\eqref{eq:transform1} and \eqref{eq:transform2}
  to model the ellipticity and orientation of the source.
\end{enumerate}

After repeated exposures with multiple grism angles the final spectrum stack
will show a superposition of source orientations.  An example of a simulated galaxy image is
shown in Fig.~\ref{fig:fastspec}.

\section{Line-interloper models}\label{app:trends}

\begin{table}
    \caption{Bypass model fit parameters.}
    \label{table:fitparams}
\centering
\begin{tabular}{lcccc}
\hline\hline
\noalign{\vskip 1pt}
Type & Redshift range &  $c$  & $\SN_0$ & $b$ \\
\hline
\noalign{\vskip 1pt}
\Ha & $0.9<z<1.5$ & 0.960 & 5.945 & 4.646 \\
\Ha & $1.5<z<1.8$ & 1.000 & 5.647 & 5.773 \\
$\rm [\ion{O}{iii}]$    & $1.5<z<1.8$ & 0.002 & 2.682 & 2.787 \\
$\rm [\ion{O}{iii}]$    & $1.8<z<2.7$ & 0.074 & 4.329 & 4.886 \\
$\rm [\ion{S}{iii}]$   & $0.3<z<0.9$ & 0.120 & 5.001 & 3.860 \\
\hline
\end{tabular}
\end{table}

\begin{figure}
    \centering
    \includegraphics[width=3.5in] {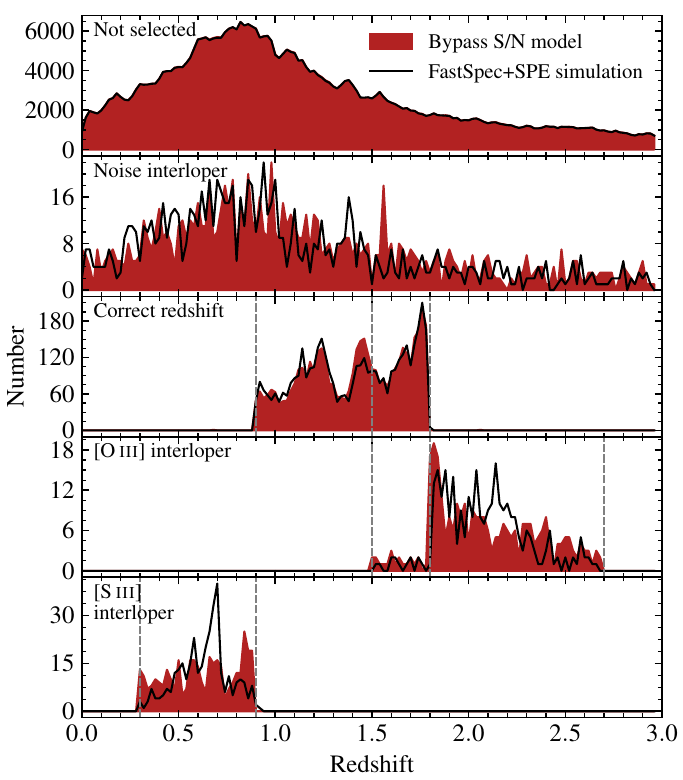}
     \caption{Predictions of line interloper redshift distributions from the bypass S/N model described in Sect.~\ref{sec:bypass} and compared to the ground truth in the EL-COSMOS \fastspec +~SPE  simulation. \label{fig:interlopernz}}
\end{figure}

\begin{figure}
    \centering
    \includegraphics[width=3.5in] {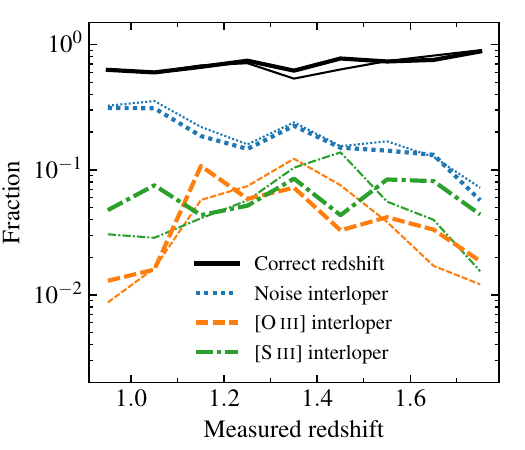}
     \caption{Predictions of line interloper redshift fractions from the bypass S/N model described in Sect.~\ref{sec:bypass} are shown in thick lines. Overplotted with thin lines are the fractions measured in the EL-COSMOS \fastspec +~SPE  simulation from Fig. \ref{fig:fractions}. \label{fig:interloperfrac}}
\end{figure}

The detection model framework described in Sect.~\ref{sec:bypass} reproduces both the input line interloper fractions and their redshift distributions by specifying only the S/N and redshift. The best-fitting detection model parameters for the \Ha, [\ion{O}{iii}], and [\ion{S}{iii}] line interlopers are given in Table~\ref{table:fitparams}. The comparisons with \fastspec+~SPE simulation ground truth are provided in Figs.~\ref{fig:interlopernz} and \ref{fig:interloperfrac}, which show very good agreement.   

\begin{figure*}
    \includegraphics[width=\linewidth] {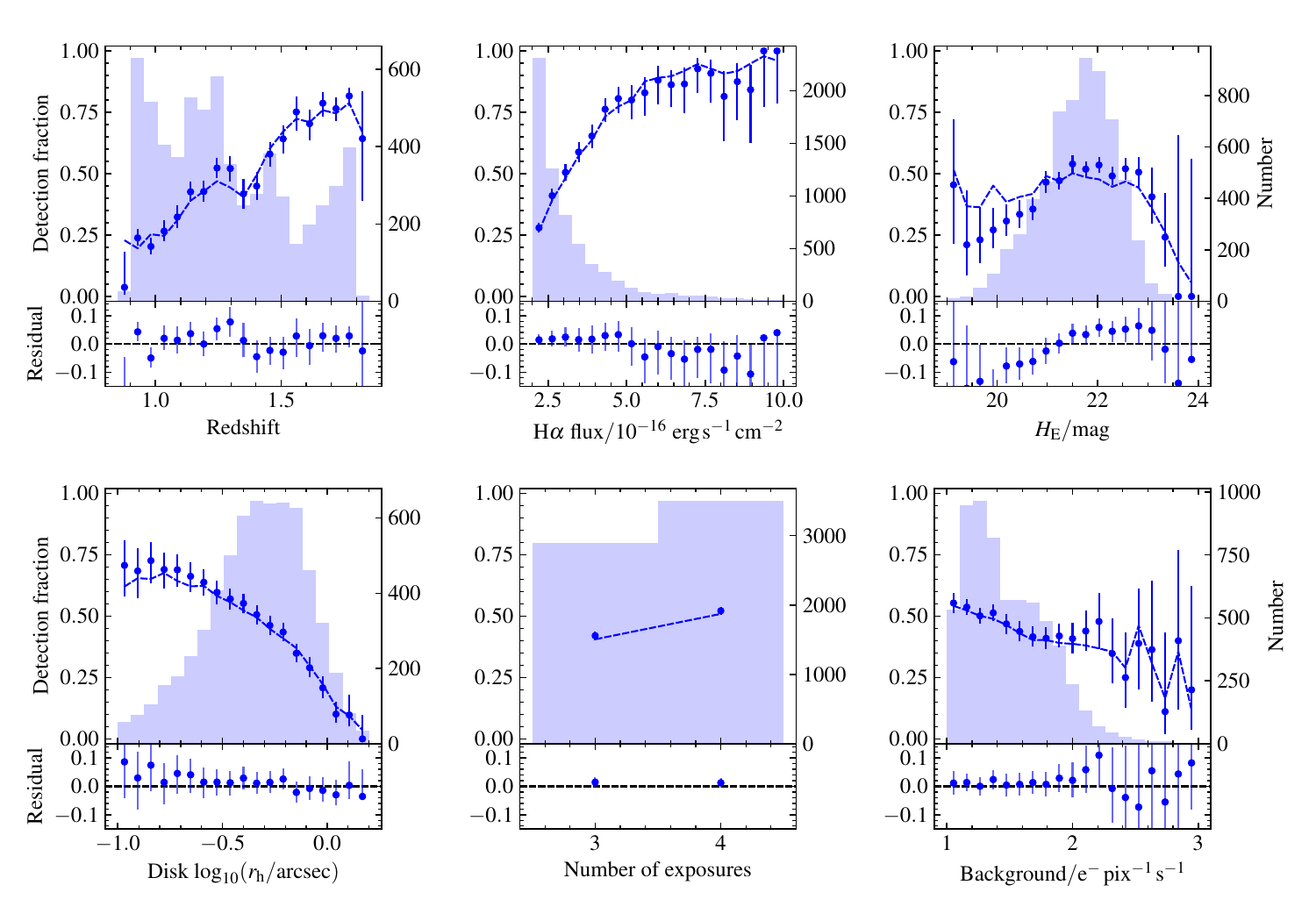}
    \caption{Trends in the detection fraction in the \fastspec+~SPE simulation as a function of galaxy and survey properties. The data points show the measured fraction in the simulation, while dashed curves are best-fitting sigmoid models that depend only on S/N and redshift. The shaded histogram shows the distribution of the sample.  \label{fig:model}}
\end{figure*}

By further examining the detection fraction related to secondary intrinsic galaxy or survey properties, we also found good agreement. This is illustrated in Fig.~\ref{fig:model}, where the detection fraction in the \fastspec+~SPE simulation is compared with the bypass S/N model predictions as a function of redshift, \Ha flux, \HE magnitude, disk scale length, exposure count, and background noise. These trends highlight the secondary dependencies that are present in the simulation due to the dependence of S/N on the galaxy properties. For example, a larger disk scale length leads to a lower S/N, and in turn to a lower detection fraction.

\end{document}